\documentclass[iop]{emulateapj}

\newcommand{\pic}[2][1]{\includegraphics[width=#1\linewidth]{#2}}

\usepackage{amsmath, amsthm, amssymb,amsfonts}
\usepackage{url}

\usepackage{tikz}
\usetikzlibrary{shapes, arrows, shadows}
\usetikzlibrary{decorations.pathreplacing}

\providecommand{\pic}[2][1]{\includegraphics[width=#1\linewidth]{pics/#2}}
\newcommand{\plotsize}{0.95}

\providecommand{\nolinkurl}[1]{\url{#1}}

\begin{document}
	
	\title{A possible advantage of telescopes with a non-circular pupil}
	
	\author{Guy Nir\altaffilmark{1}}
	\author{Barak Zackay\altaffilmark{2}}
	\author{Eran O.~Ofek\altaffilmark{1}}
	
	\altaffiltext{1}{Benoziyo Center for Astrophysics, Weizmann Institute
		of Science, 76100 Rehovot, Israel}
	\altaffiltext{2}{Institute for Advanced Study, 1 Einstein Drive, Princeton, NJ 08540, USA}
	
	\begin{abstract}
		Most telescope designs have a circular-shape aperture. 
		We demonstrate that telescopes with an elongated pupil 
		have better contrast, at lower separations, 
		between a bright central star and a faint companion. 
		We simulate images for an elongated-pupil telescope and for a circular-pupil telescope 
		of equal aperture area and integration time, 
		investigating specifically what is the maximal contrast 
		for finding faint companions around bright stars 
		as a function of angular separation.  
		We show that this design gives better contrast at lower separation from a bright star. 
		This is shown for diffraction-limited (for perfect and imperfect optics) and seeing-limited speckle images, 
		assuming equal aperture area and observing time. 
		We also show the results are robust to errors in measurement of the point spread function. 
		To compensate for the wider point spread function
		of the short axis, images should be taken at different
		rotation angles, either by rotating the telescope around the optical axis or by allowing a
		stationary mirror array to scan different parallactic angles with time. 
		Images taken at different rotation angles are added 
		using the proper image coaddition algorithms developed by Zackay \& Ofek.
		The final image has the same contrast in all angles, rather than in specific areas of diffraction nulls.
		We obtained speckle observations with a small, ground based elongated-aperture telescope 
		and show the results are consistent with simulations. 
	
	\end{abstract}
	
	
	\section{Introduction}
		
	Telescope design is usually restricted to a circular pupil, 
	which also has an approximately circular-symmetric Point Spread Function (PSF).
	In the past, telescopes with a non-circular pupil design have been considered
	as a means to improve resolution in one axis
	or to generate strong diffraction nulls that help detect faint companions or planets 
	around bright stars.
	For example, the Large Binocular Telescope \citep{Large_Binocular_Telescope_interferometric_design_Byard_1994}
	uses a non-circular pupil to achieve high resolution;
	the proposed design for \textit{DARWIN} \citep{Darwin_telescope_configurations_Kaltenegger_2006} 
	suggests to use a constellation of satellites and combine their light on large baselines, 
	which allows the detections of faint planets at very small angular separation;
	the proposed Terrestrial Planet Finder mission \citep{terrestrial_planet_finder_NASA_2003, Terrestrial_Planet_Finder_Brown_2003} 
	had two designs, one with multiple satellites and one with a coronograph. 
	The use of a non-apodized coronograph \citep{Shaped_coronograph_Spergel_2001} 
	and an aperture apodized specifically to create strong nulling of the central object 
	have been suggested for this mission \citep{Pupil_Design_for_planets_Spergel_2001}. 
	
	Here we investigate a simple, non-circular pupil design
	where information from the long axis provides higher resolution 
	(specifically, better separation of close binaries) 
	compared to a circular pupil aperture with the same pupil area. 
	The short axis of the pupil produces images of lower resolution, 
	but this is compensated for by rotating the pupil to sample the image in several angles, 
	using the coaddition algorithm
	developed in \cite{Coaddition1_Zackay_2017, Coaddition2_Zackay_2017},
	that extracts the maximal information from each spatial frequency bin.
	The rotation of the pupil, along with the coaddition method, 
	produces a final image that has high resolution in all angles, 
	and not only in specific areas of the image.
	
	For any telescope design, resolution will be hindered by 
	imperfect optics and, for ground-based telescopes, by the turbulent atmosphere. 
	We show that, under these conditions, a rotating, elongated-pupil telescope
	will outperform a symmetric design in terms of resolution. 
	We describe the image processing algorithms for adding images in multiple angles
	while conserving all the constant-in-time information in the individual images.
	The main goal of this work is to test whether telescopes with
	an asymmetric pupil design have an advantage over circular pupil telescopes
	in terms of average angular resolution and contrast.
		
	A few applications that may benefit from this approach are 
	segmented mirror telescopes, 
	where the same number of mirrors can be aligned in a row
	instead of a disk to achieve higher resolution; 
	or space-based missions, 
	where a folding telescope 
	would have less mass and moving parts
	if it only needs to open up in one axis.
	
	Some engineering constraints may complicate the construction and use 
	of elongated pupil telescopes that also need to be rotated. 
	These problems should be addressed for each individual application 
	and are beyond the scope of this work.  
	Others have recently addressed some of these issues, e.g.,
	\cite{asymmetric_scope_exoplanet_telescope_Monreal_2018} 
	have proposed a specific design implementing a telescope with a long and narrow pupil, 
	while \cite{asymmetric_scope_exoplanets_Green_2018} propose using a long and narrow aperture
	to facilitate the long baselines required for detection of exo-planets in the mid-infra red. 
	
	In \S\ref{sec: algorithm} we review the proper coaddition algorithm,
	along with some modifications, 
	and highlight the relevance to elongated-pupil telescopes.
	In \S\ref{sec: simulations} we present simulations of the elongated-pupil telescope and relevant image processing. 
	In \S\ref{sec: measurements} we present ground-based observations using a small telescope with an elongated pupil,
	and in \S\ref{sec: summary} we summarize the results and possible applications. 
	
	\pagebreak
	
	\section{Review of the image coaddition algorithm}\label{sec: algorithm}
		
	In this section, we outline the methods used to coadd images
	in the subsequent simulations and data analysis. 
	In the background-dominated case, \cite{Coaddition1_Zackay_2017, Coaddition2_Zackay_2017} 
	derived from first principles a coaddition method that is both numerically stable, 
	produces images with uncorrelated noise, 
	and preserves information at all spatial frequencies. 	
	The algorithm gives more weight to frequency bins that have more information.
	The highest frequencies in the PSF of the this coadd image
	are similar to the highest frequencies in the individual PSFs. 
	The resulting image in Fourier space is given by:
	
	\newcommand{\pj}{\widehat{P_j}}
	\newcommand{\pja}{|\widehat{P_j}|^2}
	\newcommand{\mj}{\widehat{M_j}}
	\newcommand{\mja}{|\widehat{M_j}|^2}
	\renewcommand{\dj}{\widehat{D_j}}
	\newcommand{\dja}{|\widehat{D_j}|^2}
	\newcommand{\ta}{|\widehat{T}|}
	\begin{align}\label{eq: proper coaddition full}
	\widehat{R} = \frac{\sum_j \frac{f_j}{\sigma_j^2}\overline{\pj}\mj}{\sqrt{\sum_j\frac{f_j^2}{\sigma_j^2}\pja}} \equiv \widehat{T}\widehat{P_R} + \widehat{\epsilon_R}.
	\end{align}
	Where $M_j$ are the measured images with image index $j$,
	and $P_j$ are the unity-normalized PSFs for each image. 
	The notation $\widehat{\square}$ represents a 2-dimensional Fourier transform, 
	while $\overline{\square}$ represents complex conjugation. 
	The overall flux of each image is $f_j$ and 
	the background noise standard deviation of each image is $\sigma_j$. 
	The resulting image $R$ can be represented as the true image $T$ 
	convolved with an effective PSF $P_R$, and some uncorrelated noise $\epsilon_R$. 	
	The effective PSF of the proper image is:
	\begin{align} \label{eq: proper coaddition psf}
		\widehat{P_R}=\sqrt{\sum_j \frac{f_j^2}{\sigma_j^2} \pja}.
	\end{align}

	For ground-based long-exposure images (more than a few seconds), 
	the PSF of each image can be measured by 
	observing a bright star (or multiple stars) in the frame.
	For exposures on time-scales of $\lesssim 10$\,ms, the PSF varies from image to image 
	and over small angular distances within the same image (e.g., \citealt{Isoplanatic_patch_solar_observations_Title_1975,Isoplanatic_patch_Zernike_polynomials_Chassat_1989}).
	In this case, it may still be possible to estimate the PSF using a wavefront sensor \citep{wavefront_sensor_Primot_1990, wavefront_sensor_Fried_1987}, 
	phase retrieval techniques \citep{cross_spectrum_Knox_1974, bispectrum_Lohmann_1983} or by using the image as its own PSF~\citep{Coaddition2_Zackay_2017}. 	
	
	If the PSF is known, 
	and the images are background-noise dominated, 
	the resulting image given by Equations~\ref{eq: proper coaddition full} and~\ref{eq: proper coaddition psf}
	is both optimal and a sufficient statistic\footnote{A statistic is sufficient with respect to a statistical model
		                                                and its associated unknown parameter if 
		                                                no other statistic that can be calculated from the same sample 
		                                                provides any additional information as to the value of the parameter
	                                                    \citep{foundations_statistics_Fisher_1922}.}, 
	e.g., the coadded final image can be used to find faint companions 
	or as a reference image for transient searches~\citep{Subtraction_Zackay_2016}.
	
	
	\subsection{Speckle coaddition: the square-root method}\label{subsec: cosqrt}
	
	In certain applications, 
	e.g., when obtaining speckle images
	without any nearby reference star,
	it is difficult to measure the PSF directly. 
	In this case, we can approximate the PSF of each image 
	by the image itself, as discussed in~\cite{Coaddition2_Zackay_2017}: 
	\begin{align}\label{eq: approx image psf}
		f_j P_j \approx M_j.
	\end{align}
	Here we use the approximation that the noise $\sigma_j$ is constant for all images and,
	absorb the flux term $f_j$ into the overall normalization of the PSF. 
	The resulting proper coadded image is:
	\begin{align} \label{eq: cosqrt coaddition}
		\widehat{Q} \approx \frac{\sum_j \overline{\mj}\mj}{\sqrt{\sum_j\mja}}=\sqrt{\sum_j\mja}\approx \widehat{P_R}.
	\end{align}
	The result is similar to the correlation map of the image, 
	with the critical difference that it is normalized, in Fourier space, 
	by the standard deviation of each frequency.
	This normalization ensures that 
	if the noise in the original images is independent and identically distributed (i.i.d), 
	then the noise in the resulting image is also i.i.d. 
	
	For images taken at short exposure times ($\lesssim 10$\,ms), 
	the speckle pattern of each star in every exposure 
	is a good approximation for the PSF. 
	Using Equation~\ref{eq: cosqrt coaddition} without any additional information 
	will result in an image with dramatically improved resolution compared to simply summing the images
	\citep{Coaddition2_Zackay_2017}.
	In this approximation, we do not recover the true image, $T$, 
	but an estimate of the power spectrum of the true image, $|\widehat{T}|$, 
	so that the resulting image, $Q$, is no longer a sufficient statistic 
	for general hypothesis testing or measurement. 
	However, it is still possible to answer specific statistical questions, 
	e.g., looking for faint companions around bright stars, 
	differentiating point sources from extended sources, 
	or for measuring the flux of multiple, adjacent stars. 
	
	To calculate this statistic, we zero pad each image to twice its original size, 
	Fourier transform each image and take the absolute value squared of each pixel.
	The resulting power spectrum for each image is summed 
	and, finally, the square root of the sum of power spectra is obtained. 
	When searching for point sources, the coadded image is filtered by its own PSF:
	\begin{equation}\label{eq: cosqrt filtered}
		\widehat Q \widehat{P_R} \approx \sum_j \mja. 
	\end{equation}
	Zero padding of the input images is done so that the multiplication in Fourier space
	is equivalent to convolution without cyclical boundaries (the FFT itself uses cyclical boundary conditions). 
	When performing the inverse FFT to get the coadded image, we also crop back to the original image size. 
	
	\subsection{Treatment of correlated and uncorrelated noise}\label{subsec: line noise}
	
	The image $Q$ represents a correlation map, 
	so the central pixel of the image in position space 
	contains information on the observed object but also 
	on the sum of all the noise contributions from all spatial frequencies. 
	To remove the ``zero-point" correlation of the noise,
	we subtract a constant term, the minimum of the image in Fourier plane:
	\begin{align}
		\widehat{Q}_\textbf{adj} = \widehat{Q}-\min(\widehat{Q}).
	\end{align}
	If the image suffers from strong correlated noise (e.g., line noise sometimes seen in sCMOS devices), 
	we instead subtract the minimum of each row and then the minimum of each column 
	(the order of operations is based on the strength of the line noise in the horizontal and vertical directions):
	\begin{align}
	\widehat{Q}_\textbf{adj} &= \widehat{Q}-\min_u(\widehat{Q}_{uv}) - \min_v(\widehat{Q}-\min_u(\widehat{Q}_{uv})), \text{ or }\\
	\widehat{Q}_\textbf{adj} &= \widehat{Q}-\min_v(\widehat{Q}_{uv}) - \min_u(\widehat{Q}-\min_v(\widehat{Q}_{uv})).
	\end{align}
	These treatments improve the overall image and the maximal achievable contrast. 
	Throughout this work, the algorithm presented above is used
	for simulations and measurements where no PSF data is available. 
	When PSF data is also available, 
	a different algorithm is used, as described in \S~\ref{subsec: known psf}
	
	
	\subsection{High-contrast imaging with known PSF}\label{subsec: known psf}
	
	When looking for dim companions around bright stars, 
	there is an optimal\footnote{Optimal in the sense that for a given detection rate, it has a minimal false-alarm rate.} 
	coaddition technique that is not limited to the case of background-dominated noise. 
	If the PSF of each image is known (e.g., in some space-based missions or when the wavefront aberrations are measured independently), 
	an optimal statistic can be used specifically for the detection of faint companions
	(hereafter referred to as the {\it binary coaddition} statistic). 
	This method is developed from first principles in \cite{Coaddition4_Zackay_2018} 
	and is used for all simulations where we assume the PSF is known. 
	The resulting statistic is given by:
	\begin{equation}\label{eq: binary statistical test}
		S = \sum_j f_j \left(\overleftarrow{P_j} \otimes \frac{M_j-f_jP_j}{f_jP_j+\sigma_j^2}-\sum_{x,y} P_j\frac{M_j-f_jP_j}{f_jP_j+\sigma_j^2}\right),
	\end{equation}
	and the variance for each point is given by:
	\begin{align}\label{eq: binary variance} 
		V_S &= \sum_j f_j^2 \left(\overleftarrow{P_j^2} \otimes \frac{1}{f_jP_j+\sigma_j^2}\right. \notag \\
	& \left.\quad -2\overleftarrow{P_j} \otimes \frac{P_j}{f_jP_j+\sigma_j^2}+\sum_{x,y} \frac{P_j^2}{f_jP_i+\sigma_j^2}\right).
	\end{align}
	Here we use $\overleftarrow{\square}$ to denote coordinate reversal ($x,y\to-x, -y$)
	and $\otimes$ to denote 2-dimensional convolution. 
	Using this statistic, which is optimal for the detection of high-contrast companions, 
	the detection limit for each point in the image is given by:
	\begin{equation}\label{eq: binary contrast}
		C = \frac{\sqrt{V_S}}{S/N},
	\end{equation}
	where $S/N$ is the signal to noise ratio (in units of standard deviation) required for detection. 
	In this work, we adopt $S/N=5$ as a threshold for detection. 
		
		
	\subsection{Coaddition of elongated-pupil telescope images}\label{subsec: coadding asymmetric}
		
	In the case of the elongated-pupil telescope,
	the images have a wide PSF in one direction, 
	corresponding to the narrow axis of the pupil. 
	By imaging the source at different rotation angles 
	(i.e., different position angles of the pupil's long axis as projected on the sky), 
	and using the algorithm of \cite{Coaddition2_Zackay_2017} and presented in \S\ref{subsec: known psf}, 
	information is recovered from frequency bins in all directions,
	and the resulting image will have a resolution that is as good as 
	a telescope with a full, circular aperture of the same diameter 
	as the long axis of the elongated-pupil telescope. 
	
	In order to recover a final image that has a symmetric, round PSF, 
	the pupil needs to be rotated around the optical axis by a full 180 degrees, 
	and the position angle, $\theta$, should be sampled in intervals smaller than 
	the ratio between the long and narrow sides of the PSF. 
	In our simulation (with a width-length ratio of 10), 
	we found that $\Delta\theta \sim 5^\circ$ is sufficient (see \S~\ref{subsec: rotation angles}). 
	
	
	\section{Simulations}\label{sec: simulations}
	
	We conducted several sets of simulations to compare the theoretical ability 
	of circular- vs.~elongated-pupil telescopes to detect
	close and faint companions around bright stars. 
	We simulated PSFs from a circular and an elongated pupil. 
	A 180\,cm diameter was chosen for the circular pupil,
	while the elongated-pupil telescope had a 500 by 50\,cm, rectangular aperture.
	This results in both telescopes having nearly the same aperture area. 
	
	The goal of these simulations was to test the relative performance of the
	circular- and elongated-pupil telescopes, 
	not the coaddition algorithms themselves. 
	The same methods were used for both telescopes, 
	so they could be compared under equivalent conditions. 
	
	We ran three sets of simulations, 
	the first assuming perfect optics (\S~\ref{subsec: simulations perfect});
	the second assuming imperfect optics (\S~\ref{subsec: simulations red noise}), 
	where the wavefront aberrations have ``red-noise'' properties\footnote{In the case of optics, the aberrations typically have a higher amplitude at lower spatial frequencies in the pupil plane.}; 
	and the last set using atmospheric aberrations of the wavefront (\S~\ref{subsec: simulations atmosphere}).
		
	The code used is available as part of the MATLAB astronomy \& astrophysics toolbox~\citep{matlab_package_Ofek_2014}. 
	
	\subsection{Simulation of diffraction limited images}\label{subsec: simulations perfect}
	
	In the first set of simulations, 
	we assume the optics are perfect and the PSF is known.
	This regime represents the absolute maximal contrast attainable 
	for either circular- or elongated-pupil telescopes.
	
\begin{figure}
		\centering
		\pic[1]{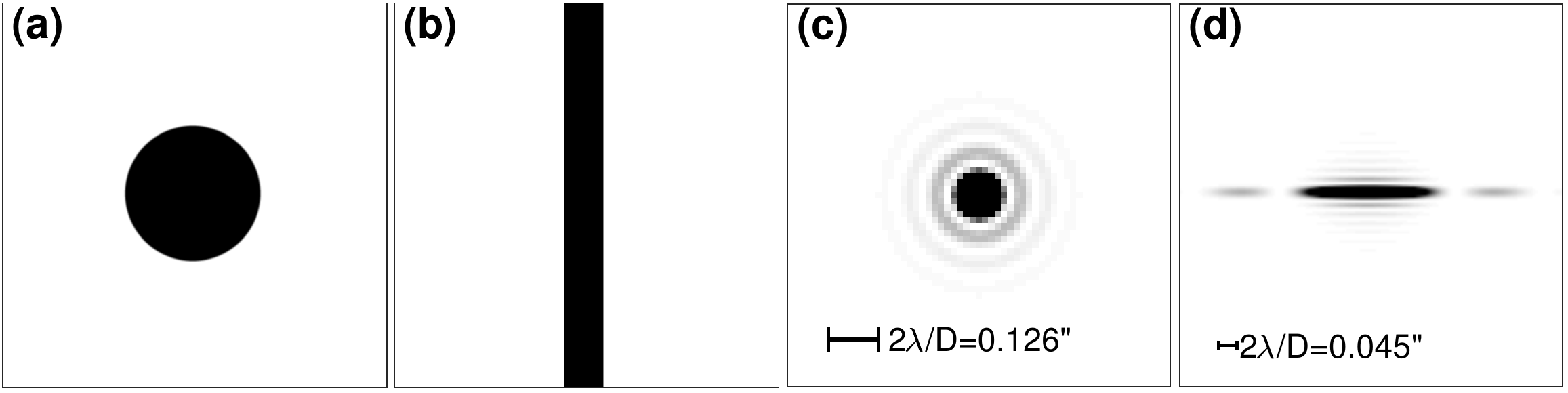}
		\caption{Simulated pupils and PSFs for the circular- and elongated-pupil telescopes.
			(a) The aperture shape of the circular-pupil telescope has a shorter diameter than the long edge of the elongated-pupil telescope.
			(b) The aperture shape of the elongated-pupil telescope has the same area as the circular pupil. 
			(c) The PSF from the circular pupil is an Airy disk. 
			(d) The PSF from the elongated pupil is wider on one axis but narrower on the other. 
			The two PSFs are shown on the same angular scale, with the diffraction limit shown in each corner for reference. 
			}
		\label{fig: psf examples perfect}
	\end{figure}

	We simulated a 10th magnitude star observed in the $V$ band 
	for 36 exposures of 125 seconds, 
	equivalent to a total of approximately $10^{10}$ photons. 
	A circular aperture was employed to simulate a symmetric PSF, 
	which is used in two simulations: 
	the {\it un-rotated simulation}, where a single orientation is used 
	for a long exposure of $36\times 125$ seconds;
	and the {\it rotated simulation} where the PSF was rotated by 5 degrees 
	between each of 36 consecutive orientations, exposing each for 125 seconds. 
	The elongated-pupil aperture was used to produce an elongated PSF, 
	which was used in 36 consecutive 125-second exposures with a 5 degree rotation angle step. 
	To minimize interpolation errors, 
	we used three-shear rotation, 
	using Fourier interpolation for the different skews~\citep{Fourier_rotations_Larkin_1997}.
	For each exposure, the pixel scale was set to 32 pixels per $\lambda/D$, 
    where $D$ was chosen to be the long axis of each telescope. 
    Oversampling by a factor of 16 over the Nyquist sampling was used so that features (such as the contrast curves) 
    of the resulting coadded images could be measured more accurately. 
    In real measurements, a sampling of 2 pixels per $\lambda/D$ is sufficient. 
	Source noise and an additional noise of 1 electron/pixel (e.g., read noise) were added to all images. 
	The noise for the image of the un-rotated circular-pupil simulation 
	was set to $\sqrt{36}$ to compensate for having one exposure instead of many. 
	The pupils and resulting PSFs for the circular- and elongated-pupil simulated telescopes are shown in Figure~\ref{fig: psf examples perfect}
	
\begin{figure}
		\centering
		\pic[\plotsize]{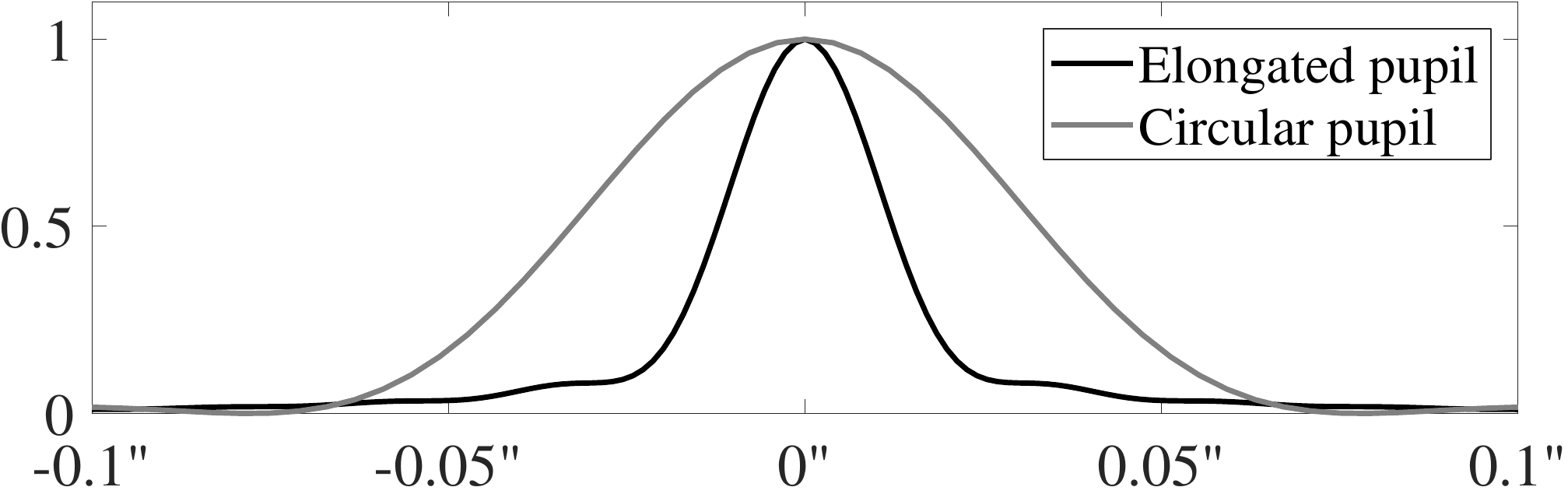}
		\caption{A 1D profile (cut through) of the proper coaddition PSF, ${P_R}$, 
			as given by Equation~\ref{eq: proper coaddition psf}. 
			The PSF of the elongated pupil (black line), 
			is narrower than that of the circular pupil (grey line).}
		\label{fig: profile examples}
	\end{figure}

\begin{figure}
		\centering
		\pic[\plotsize]{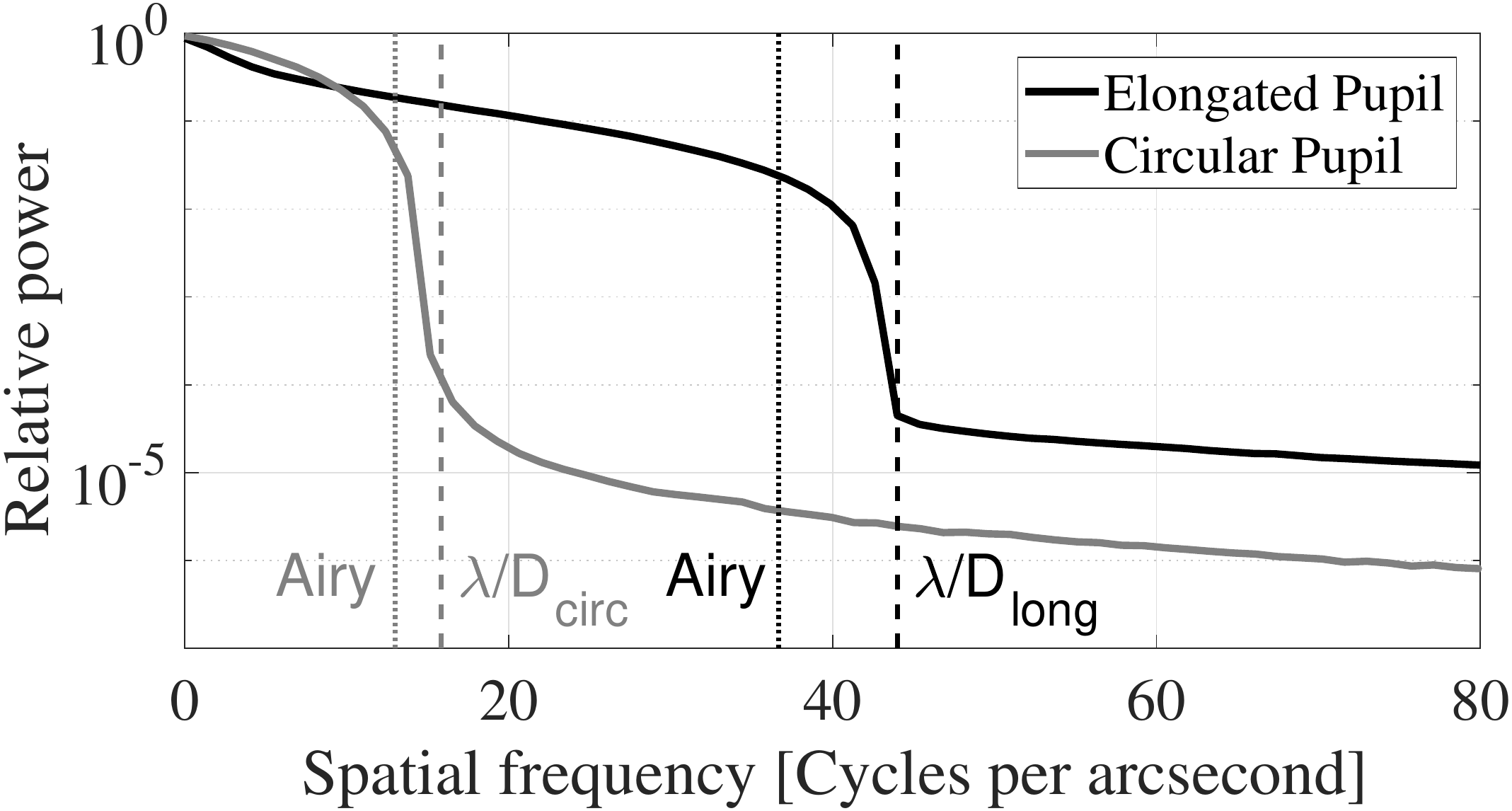}
		\caption{A 1D profile (integrated over circles of increasing radii) of the MTF of the two apertures for the case of perfect optics. 
			The MTF is given as the absolute value of the proper coaddition PSF, in Fourier space, $|\widehat{P_R}|$.
			The elongated-pupil MTF has information content at higher frequencies than the circular-pupil MTF. 
			The theoretical diffraction limit of the circular- and elongated-pupils is shown as gray and black dashed lines, respectively.
			The radius of the inner circle of the Airy disk ($1.22\lambda/D$) for each pupil is given by the gray and black dotted lines. 
			For the elongated pupil we see that there is information in the MTF up to the diffraction limit calculated based on the long edge of the pupil.}
		\label{fig: mtf perfect}
	\end{figure}

	The images were summed using the generalized proper coaddition algorithm given in Equation~\ref{eq: proper coaddition full}.
	In Figure~\ref{fig: profile examples}, we show, for both apertures, a 1D cut through the resulting PSF, $P_R$ from Equation~\ref{eq: proper coaddition psf}.
	In Figure~\ref{fig: mtf perfect}, we show the 1D Modulation Transfer Function (MTF), 
	which is the absolute value of the Fourier transform of the coadded PSF, $|\widehat{P_R}|$.
	
\begin{figure}
			\centering
			\pic[\plotsize]{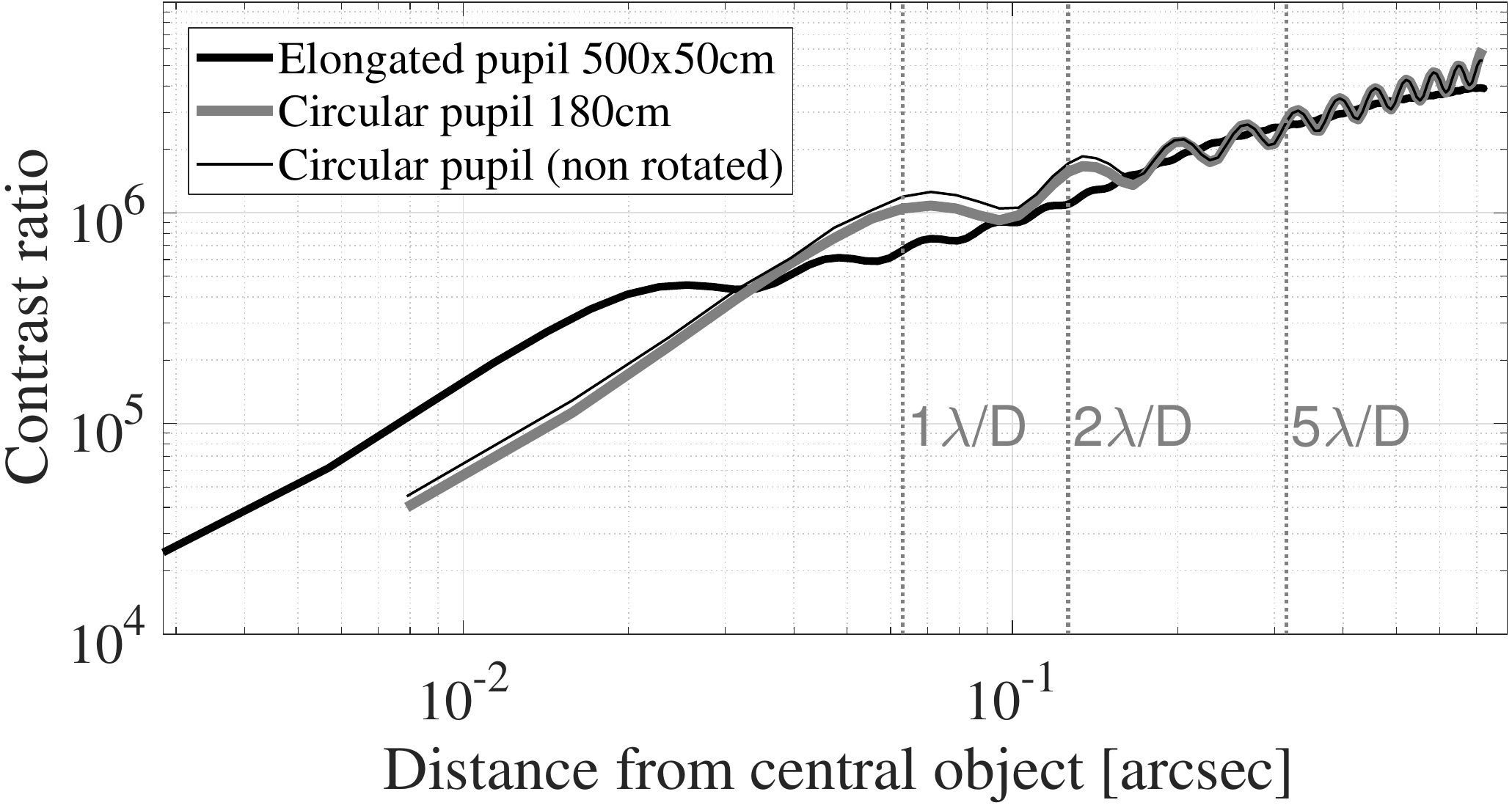}
			\caption{The highest contrast between a primary star and a companion
					 that can be detected at $S/N=5$, for different angular separations.
					 This simulation of circular and elongated pupil telescopes
					 used perfect optics and no atmospheric aberrations. 
				     The bottom $x$-axis uses angular scale for comparing different pupil sizes, 
					 but can be scaled linearly if the sizes of all pupils are changed by the same proportion. 
					 The upper $x$-axis shows the angular distance in units of the diffraction limit of the circular pupil.
					 All simulations used the same total exposure time and aperture area.
				     The thin, black curve describes the result for a single orientation of a circular telescope with an aperture of 180\,cm.
				     The thick, grey curve describes the results for 36 images taken at different angles, with a circular telescope. 
				     The thick, black curve describes the results for 36 images taken at different angles using a $50\times 500$\,cm elongated-pupil telescope.
				     As expected for perfect optics, the rotated and un-rotated circular telescopes give nearly identical results. 
					 The contrast is similar for the circular- and elongated-pupil telescopes
					 until reaching the diffraction limit, at which point the contrast drops sharply.
					 The elongated pupil has a smaller diffraction limit, consistent with the long axis of the pupil 
					 and, thus, preserves contrast at lower angular separations. 
			}
			\label{fig: contrast perfect}
		\end{figure}

	The high-contrast statistic given by Equations~\ref{eq: binary statistical test} and \ref{eq: binary variance} 
	was used to calculate the maximum contrast for detection of a faint companion around the central star. 
	We used the map of the square root of the variance as in Equation~\ref{eq: binary contrast}, 
	and calculated the median of all the pixels at a certain radius from the center,
	i.e., within an annulus with a two-pixel width. 
	The circular-pupil simulations resulted in a variance map with many features, 
	similar to the rings in an Airy disk,
	so that each annulus contained pixels with widely varying contrast values;
	a median statistic for each annulus was found to give a stable contrast values.
	To find the contrast, we assumed a $5\sigma$ detection threshold. 
	The results for perfect optics in the known-PSF case are shown in Figure~\ref{fig: contrast perfect}.
	
	The simulations show that the elongated-pupil telescope 
	has similar contrast to that of a circular telescope with the same area, 
	as long as the separation is larger than the diffraction limit of the circular telescope. 
	Inspecting the contrast curves below $0.06''$ ($\lambda/D$ of the circular telescope)
	or the MTF in Figure~\ref{fig: mtf perfect}, 
	we can see that the elongated pupil preserves information 
	on smaller angular scales than the circular pupil. 
	In fact, the simulations suggest that the theoretical resolution of the elongated pupil 
	is the diffraction limit of the long axis. 
	For a 1:10 axis ratio, the diffraction limit is 2.78 times smaller than for a circular pupil of the same area. 
	
	Note that the contrast curve of the rotated and non-rotated circular telescope are nearly identical. 
	While this is expected, since the circular-pupil PSF has circular symmetry, 
	it also shows that proper rotation (using FFT-skew transformations) does not bias the results. 
	
	All angular scales are shown for specific aperture sizes that were chosen arbitrarily, 
	and can be scaled linearly to any telescope size. 
	For example, the contrast at 0.1 arcsecond for the 180\,cm circular pupil (or $500\times 50$\,cm elongated pupil) 
	is equivalent to the same contrast at 0.05 arcsecond for a 360\,cm (or $1000\times 100$\,cm) telescope, 
	as long as the total photon count is preserved. 
	
	\subsection{Simulation of diffraction limited images with non-perfect optics}\label{subsec: simulations red noise}
	
	Real-life telescopes have imperfect optics.
	Simulations were made for apertures with imperfections of the optical surfaces. 
	The amplitude of the deviation of the aberrated wavefront from the planar wavefront 
 	was randomly chosen from a normal distribution with a variance following a power law in the spatial frequency:
	\begin{equation}\label{eq: red noise power law}
		\sigma_k^2 = A f^{-2},
	\end{equation}
	where $A=0.2$\,rad is the RMS deviation of the base frequency, 
	and $f$ is the spatial frequency in units of pixel$^{-1}$.
	The base frequency is set by the size of each of the pupils (the diameter or the long edge). 
	The aberrated PSFs are then rotated and used to produce images (as was done in \S~\ref{subsec: simulations perfect}). 
	The images were coadded using proper coaddition, as in Equation~\ref{eq: proper coaddition full}.
	The resulting PSF and MTF showed no substantial difference from the perfect optics case. 
	This is because the inputs to the coaddition algorithm are the aberrated PSFs, 
	so that the imperfections are accounted for in the weighing of the coadded image. 

\begin{figure}
		\centering
		\pic[\plotsize]{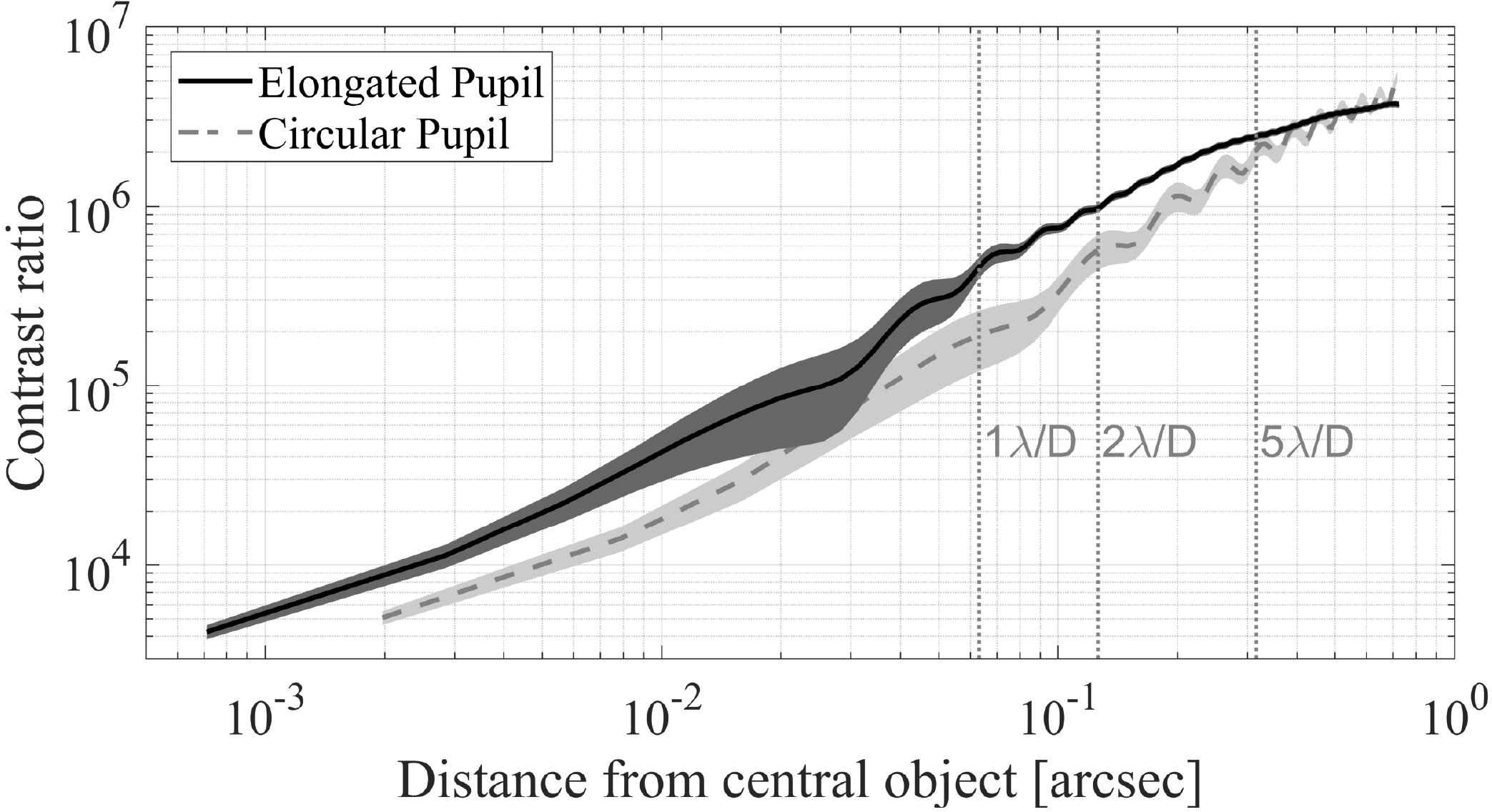}
		\caption{The same as Figure~\ref{fig: contrast perfect} but with realistic aberrated (red-noise) optics. 
				 No atmosphere is used and knowledge of the PSF is assumed. 
				 To avoid strong biases from individual realizations of the optical aberrations, 
				 we performed 30 simulations identical to those described in \S\ref{subsec: simulations perfect}.
				 The mean of the results is shown with a shaded area indicating the $1\sigma$ scatter. 
				 As expected with realistic optics, both telescopes begin to lose contrast 
				 at separation angles greater than the diffraction limit. 
				 The elongated pupil is more robust to optical aberrations, 
				 Likely because it samples only a narrow band 
				 out of the space of possible spatial frequencies of the aberrations. 
		}
		\label{fig: contrast red noise}
	\end{figure}

	The images were also coadded using Equations~\ref{eq: binary statistical test} 
	and \ref{eq: binary variance} (i.e., assuming the PSF is known). 
	The resulting contrast curves for the rotated circular-pupil telescope
	and the elongated-pupil telescope are shown in Figure~\ref{fig: contrast red noise}.	
	As each realization of the optical aberrations had a very different contrast curve,
	we avoided strong biases that may arise from any specific realization by simulating 30 complete sets 
	of 36 images of each pupil shape. 
	The resulting contrast curves were averaged, 
	and the variation between each simulation was used to derive $1\sigma$ scatter intervals.

	
	
	As expected, the addition of optical imperfections reduces the contrast achieved in all the simulations. 
	This causes both telescopes to start losing contrast at higher separations than the diffraction limit, 
	but the elongated telescope is more robust to the aberrations. 
	This is likely due to the fact that the elongated pupil samples only 
	a narrow band out of the space of possible spatial frequencies of the aberrations. 
	
	\subsection{Simulation of speckle images}\label{subsec: simulations atmosphere}
	
	To simulate ground-based telescope observations, 
	we generated a random phase-screen using the prescription in~\cite{phase_screen_simulation_Jia_2015}. 
	We gave each spatial frequency a random value from a normal distribution following the power law:
	\begin{equation}
		\sigma^2(f) = 0.033 \left(\frac{r_0}{D}\right)^{-5/3} (f_o^2+f^2)^{-11/6} \exp(-f^2/f_i^2),
	\end{equation}
	where $f=|\vec{f}\, |$, is the size of the 2-dimensional 
	spatial frequency vector $\vec{f}$, in units of pixel$^{-1}$, 
	while $f_i=5\times 10^3$\,pixel$^{-1}$ and $f_o=0.5$\,pixel$^{-1}$ 
	are the frequencies corresponding to the inner and outer scale of the atmosphere, respectively\footnote{
		The outer scale is the scale at which energy enters the atmosphere, 
		from which it cascades in turbulent eddies down to smaller scales 
		until reaching the inner scale, where dissipation overtakes turbulence. 
		Here we assume the outer scale is 10m, while the inner scale is 1mm. 
	}.
	The Fried length is given by $r_0=10$\,cm, corresponding to a seeing of $\sim 1''$, 
	and the telescope aperture is given by $D$. 
	For each image, we produced one phase screen wide enough to encompass 
	both the circular- and elongated-pupil mask arrays, 
	so that both telescopes were simulated to view 
	a different cut out of the exact same sky in each of the images.
	The resulting PSFs were used for three simulations: 
	the circular PSF with source noise and an additional noise of 1 count/pixel; 
	the elongated PSF with source noise and an additional noise of 1 count/pixel; 
	and the same elongated PSF with noise scaled by the angular size of each pixel 
	(e.g., scaling the noise as background rather than readout noise). 
	
	The number of photons in the simulations was set to be equivalent 
	to that produced by a magnitude 4 star in the $V$ band. 
	Batches of 50 images were simulated at a 10\,ms exposure time for each image. 
	For each batch, the PSF was rotated by 5 degrees, with 36 batches covering the range of 180 degrees. 
	The high magnitude was chosen so that the total photons in all exposures is $\sim 10^{10}$, 
	while requiring only a few hundred simulated images.
	
\begin{figure}
		\centering
		\pic[1]{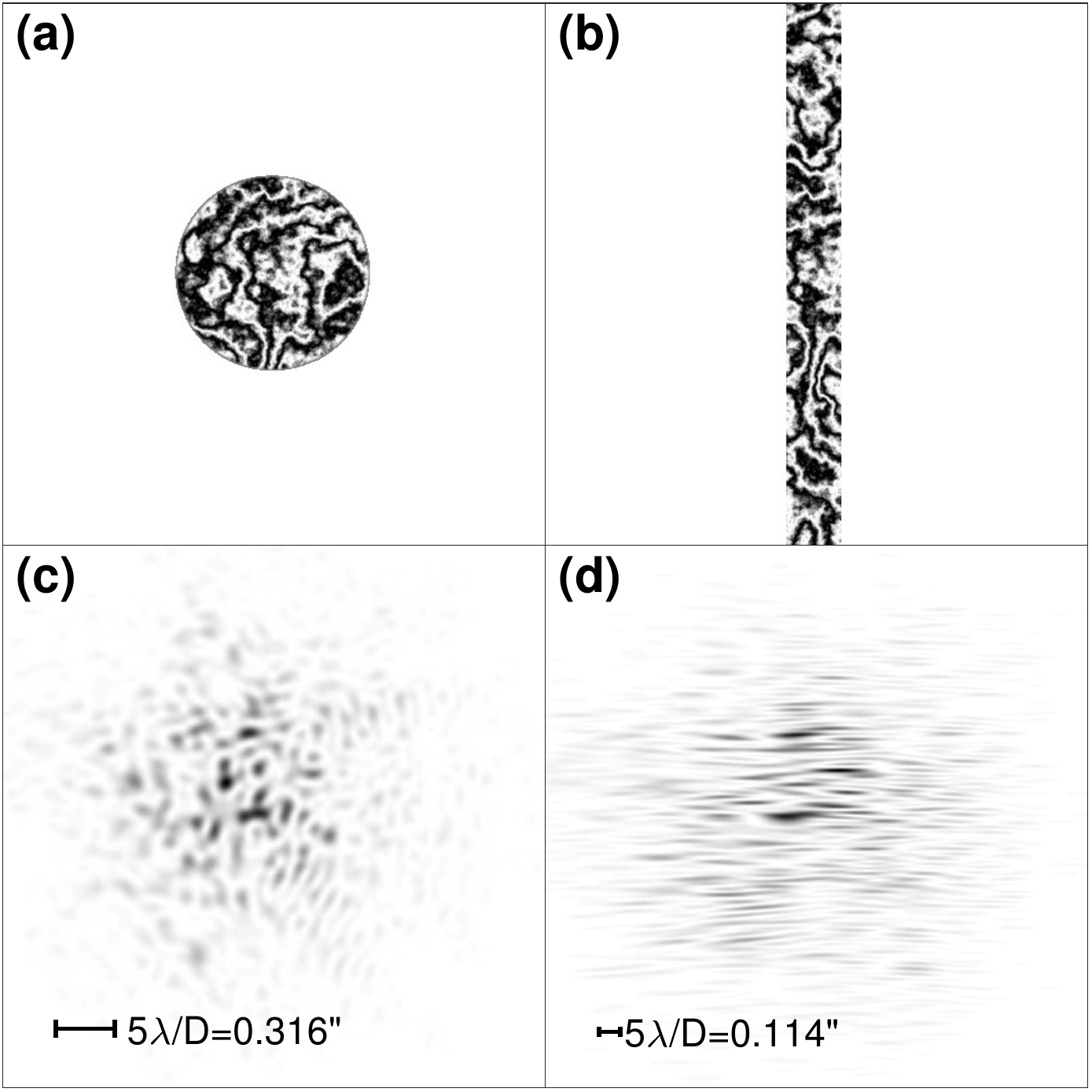}
		\caption{Example pupils and PSFs for atmospheric simulations. 
			(a) Circular-pupil telescope overlaid with the phase screen. 
			(b) Elongated-pupil telescope overlaid with the same phase screen.
			(c) Resulting PSF for the circular pupil, showing circular speckles. 
			(d) Resulting PSF for the elongated pupil, showing long and narrow speckles. 
			Both speckle patterns are shown on the same angular scale, 
			with the diffraction limit displayed for reference. 			
			}
		\label{fig: psf examples atmosphere}
	\end{figure}

	The pupils for the two telescopes, overlaid with an example phase screen, 
	are presented in Figure~\ref{fig: psf examples atmosphere}, 
	along with the resulting speckle PSFs.
	
	We used the same simulation to compare the two telescopes using three coaddition methods: 
	the proper coaddition method given by Equation~\ref{eq: proper coaddition full}, 
	the approximation given by Equation~\ref{eq: cosqrt coaddition} for situations where the PSF is not known,
	and the specialized, high-contrast imaging method given by Equation~\ref{eq: binary statistical test}, 
	where we assume the PSF is known.
	
\begin{figure}
		\centering
		\pic[\plotsize]{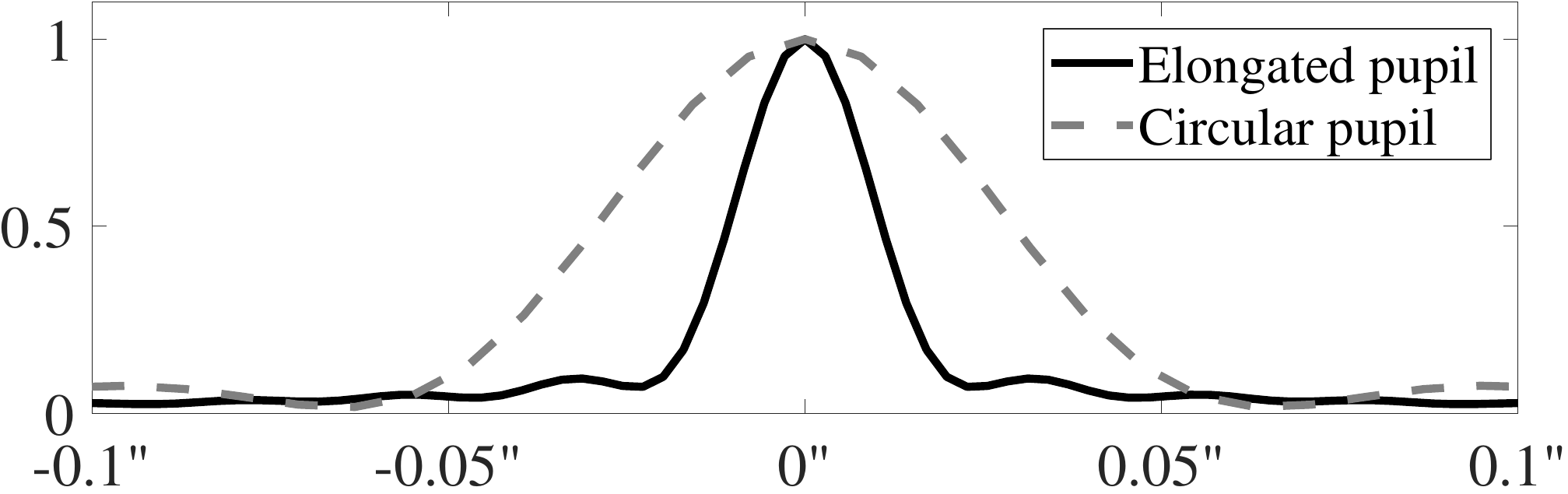}
		\caption{A 1D profile (cut) through the proper coaddition result's PSF, ${P_R}$, 
			as given by Equation~\ref{eq: proper coaddition psf}, 
			for simulations under atmospheric conditions of $r_0=10$\,cm, assuming knowledge of the PSF. 
			The PSF of the elongated pupil is narrower than that of the circular pupil.}
		\label{fig: profiles atmosphere}
	\end{figure}

\begin{figure}
		\centering
		\pic[\plotsize]{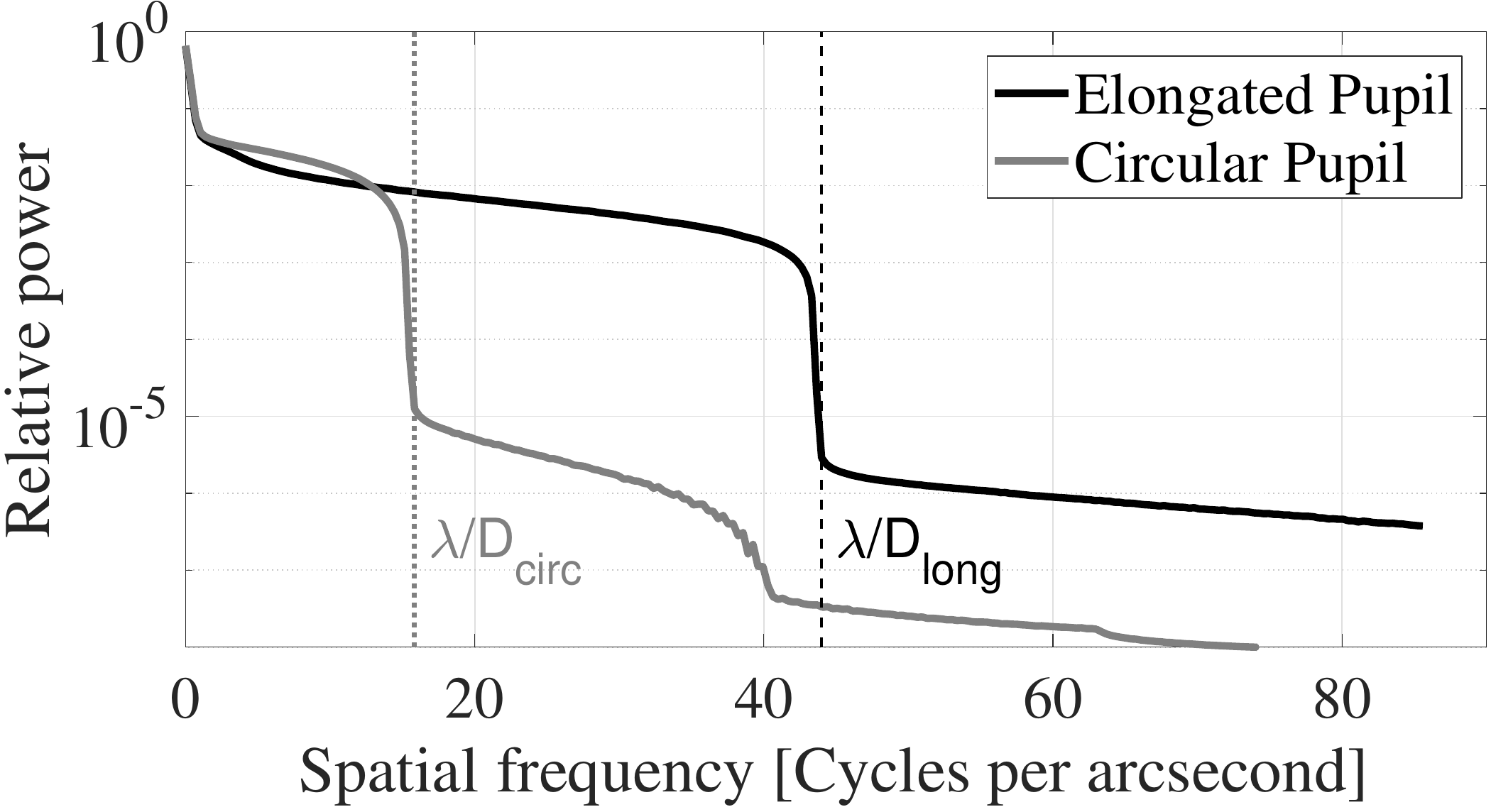}
		\caption{A 1D profile (integrated over circles of increasing radii) 
			through the MTF of the two apertures,
			under atmospheric conditions with Fried length of $r_0=10$\,cm, assuming knowledge of the PSF.  
			The elongated-pupil MTF has information content at higher frequencies than the circular-pupil MTF. 
			The theoretical diffraction limit of the circular- and elongated-pupils is shown as dotted and dashed lines. 
			For the elongated pupil, we see that there is information in the MTF up to the diffraction limit, 
			calculated based on the long edge of the pupil.}
		\label{fig: mtf atmosphere}
	\end{figure}

	The coadded PSF, $P_R$, from Equation~\ref{eq: proper coaddition psf} is shown in Figure~\ref{fig: profiles atmosphere}, 
	while the MTF, $|\widehat{P_R}|$ is shown in Figure~\ref{fig: mtf atmosphere}. 
	As before, the elongated pupil gives a narrower PSF, 
	and preserves information on higher spatial frequencies than the circular pupil. 

\begin{figure}
		\centering
		\pic[\plotsize]{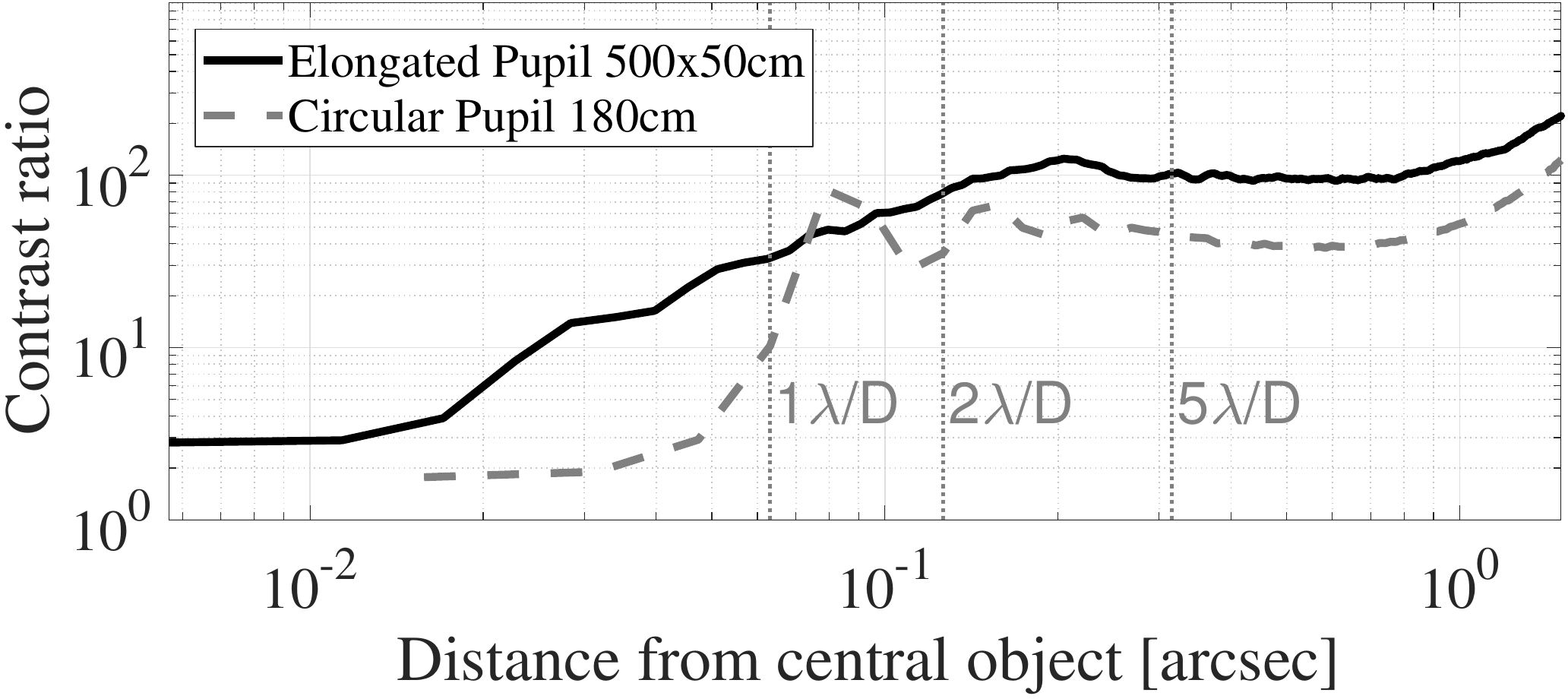}
		\caption{The same as Figure~\ref{fig: contrast perfect}, 
			but with the addition of the effects of atmospheric turbulence with a Fried length of $r_0=10$\,cm, 
			while observing in the V band, 
			corresponding to $\approx 1''$ seeing. 
			No knowledge of the PSF is assumed, 
			so images were coadded using Equation~\ref{eq: cosqrt coaddition}.
			The solid black curve describes a circular-pupil telescope with an aperture of 180\,cm, 
			The dashed grey curve describes an elongated-pupil telescope with an aperture of $50\times 500$\,cm.
			The elongated-pupil telescope has better contrast, especially in the region below 
			the diffraction limit of the circular telescope, denoted by the left-most dotted grey line.
			The annotations of multiples of the diffraction limit $\lambda/D$ are for the circular aperture diameter. 
			The results are given for specific telescope sizes and cannot be rescaled easily, 
			unless the Fried length is also scaled. 
		}
		\label{fig: contrast atmosphere cosqrt}
	\end{figure}

	The contrast curves for the circular- and elongated-pupil telescopes for the case of unknown PSF are presented in Figure~\ref{fig: contrast atmosphere cosqrt}. 
	To find the contrast in this method, 
	we compared the peak intensity to the noise in an annulus with a 2 pixel width 
	in the filtered coadded image, as given by Equation~\ref{eq: cosqrt filtered}. 
	Because there is no phase information in this method, 
	the intensity of the companion would be split symmetrically in to two peaks 
	on either side of the center of the image. 
	So we divide the contrast we find by a factor of two and by another factor of $N_\sigma=5$ 
	to find the contrast at which we expect to make a $S/N=5$ detection.
	
	The simulations show that the elongated-pupil telescope, for the same area, 
	has similar contrast to that of the circular telescope 
	down to the diffraction limit of the circular-pupil telescope.
	The elongated-pupil telescope preserves contrast at smaller separation angles, 
	down to the diffraction limit set by the long axis of the pupil. 
	Compared to the previous results for diffraction limited images, 
	the contrast for both apertures is lower, 
	due to the atmospheric conditions and lack of knowledge of the PSF. 

\begin{figure}
		\centering
		\pic[\plotsize]{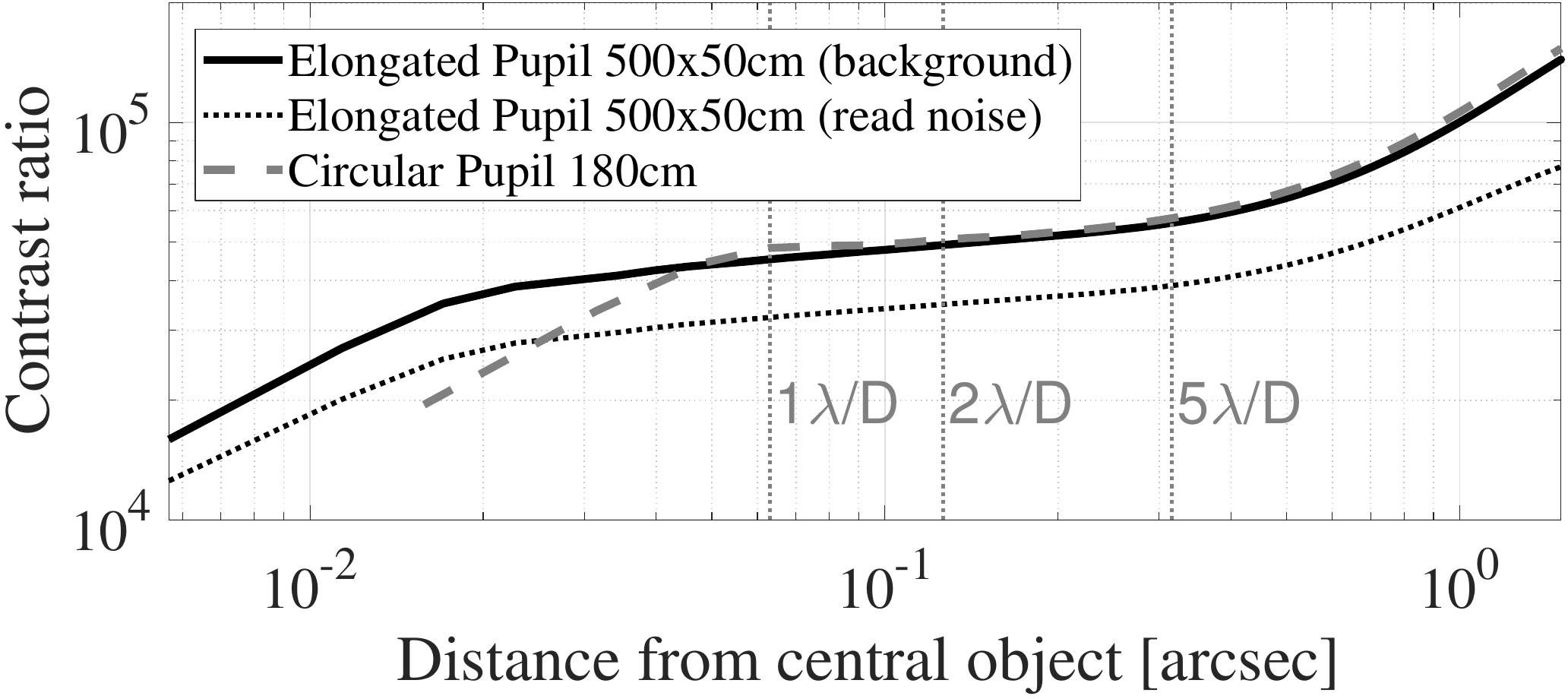}
		\caption{The same as Figure~\ref{fig: contrast atmosphere cosqrt}, 
				 but assuming knowledge of the PSF, 
			     using the coaddition method defined in Equation~\ref{eq: binary statistical test}.
			     The results of this coaddition method give substantially better contrast, 
			     emphasizing the importance of measuring the PSF along with the speckle images. 
			     The dotted black line represents the elongated pupil 
			     with noise of equal magnitude as the circular pupil. 
			     The results using equal noise show lower contrast, 
			     since the elongated-pupil simulation has smaller pixels 
			     and, thus, more affected by the noise.
			     The solid black line represents the elongated pupil 
			     with noise scaled by the angular size of pixels 
			     (i.e., sky background rather than read noise).
			     The curves for the scaled-noise elongated pupil 
			     and the circular pupil have nearly the same contrast 
			     above the diffraction limit of the circular telescope, 
			     suggesting that the shape of the individual speckles
			     has little effect on the resulting contrast. 
			     Below the diffraction limit of the circular telescope, 
			     denoted by the left-most dotted grey line, 
			     the contrast of the elongated pupil is higher, 
			     since the diffraction limit of the elongated pupil is smaller
			     when the area of the two telescopes is the same. 
		}
		\label{fig: contrast atmosphere binary}
	\end{figure}

	In the second case, 
	we assume knowledge of the PSF for each speckle image, 
	e.g., by using a wave-front sensor.
	In this case we use the method for detecting high-contrast companions, 
	given by Equation~\ref{eq: binary statistical test}.
	The results are shown in Figure~\ref{fig: contrast atmosphere binary}.
	The contrast is dramatically improved in comparison to the previous method, 
	but the atmosphere still reduces the attainable maximum contrast 
	as compared to the diffraction-limited case. 
	For this coaddition method, we show two simulations of the elongated pupil, 
	one with noise of equal magnitude as the circular-pupil telescope, 
	represented by the dotted black line, 
	and one with noise scaled by the angular size of the pixels, 
	represented by the solid black line. 
	The elongated pupil has smaller pixels to match the smaller diffraction limit of the telescope. 
	If the noise is constant per pixel (e.g., read noise), 
	the contrast from the elongated pupil is lower than from the circular pupil. 
	If the noise is constant per angular area (e.g., sky background) the contrast for the two pupil shapes
	is very similar for angular separation greater than the diffraction limit of the circular telescope. 
	Below that limit, the circular-telescope contrast is lower than the elongated-telescope contrast. 
	Since the diffraction limit of the elongated telescope is set by the long edge of the pupil, 
	its contrast begins to diminish at an angular separation that is 
	about three times smaller than the limit for the circular telescope. 
	
	\pagebreak
	
	\subsection{The effect of uncertainty in the PSF}\label{subsec: psf errors}
	
	The binary coaddition method (Equation~\ref{eq: binary statistical test}) depends on knowledge of the PSF. 
	We tested the ability of this method to recover a faint companion around a bright star 
	in situations where the PSF is not perfectly measured. 
	Contrast curves are calculated analytically using the input PSFs, and not the images, as seen in Equation~\ref{eq: binary variance}. 
	Therefore, we cannot use them to test the effects of the difference between the aberrated PSFs
	and the images, which are simulated using the original PSFs. 
	Instead, we planted a faint companion with a contrast ratio of 1:$10^4$
	at an angular separation of $0.5''$ from a simulated 4-th magnitude star. 
	We simulated atmospheric conditions and used perfect optics. 
	We generated 18,000 images, in batches of 100 images separated by an angle of 1 degree between each batch. 
	We used the same atmospheric phase screen for both the circular- and elongated-pupil telescope simulations, 
	and coadded the resulting images. 
	The PSFs produced by the atmospheric phase screen were used directly to create the images 
    by multiplying by the flux of the two stars, with the correct offset for the faint companion, 
    and then adding noise proportional to the angular scale of each pixel, 
    i.e., 1 count per pixel for the circular-pupil telescope and 0.13 counts per pixel for the elongated-pupil telescope.
    The PSFs were given to the coaddition algorithm only after adding Gaussian noise to each pixel of the PSF, 
    with RMS proportional to the intensity in that pixel.     
    We repeated this simulation 5 times and averaged the score maps of each coaddition result. 
    
\begin{figure}
    	\centering
    	\pic[1]{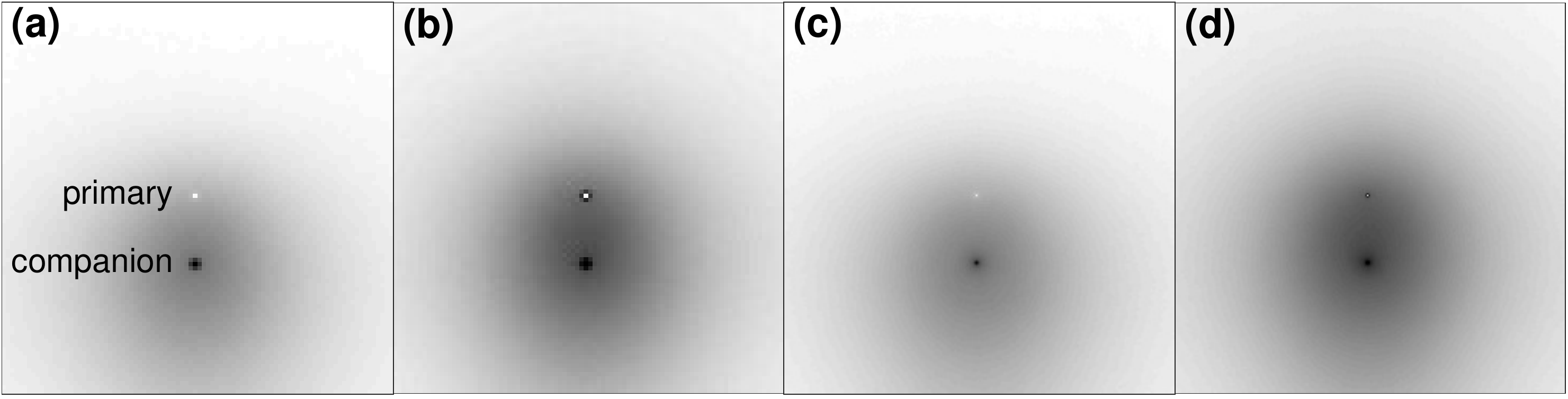}
    	\caption{Example images of the statistical score map for the binary coaddition method given in Equation~\ref{eq: binary statistical test} 
    		for circular- and elongated-pupil telescope simulations with a companion at a contrast ratio of 1:$10^4$ at an angular separation of 0.5$''$. 
    		(a) The circular-pupil score map with perfect PSF measurement. 
    		(b) The circular-pupil score map with Gaussian random PSF errors with RMS of 2\% of each PSF pixel value. 
    		(c) The elongated-pupil score map with perfect PSF measurement. 
    		(d) The elongated-pupil score map with Gaussian random PSF errors with RMS of 2\% of each PSF pixel value.  
    		The errors in the PSF cause an increased ``background" around the central star thereby washing out the signal from the companion. 
    	}
    	\label{fig: examples psf errors frac}
    \end{figure}

\begin{figure}
		\centering
		\pic[1]{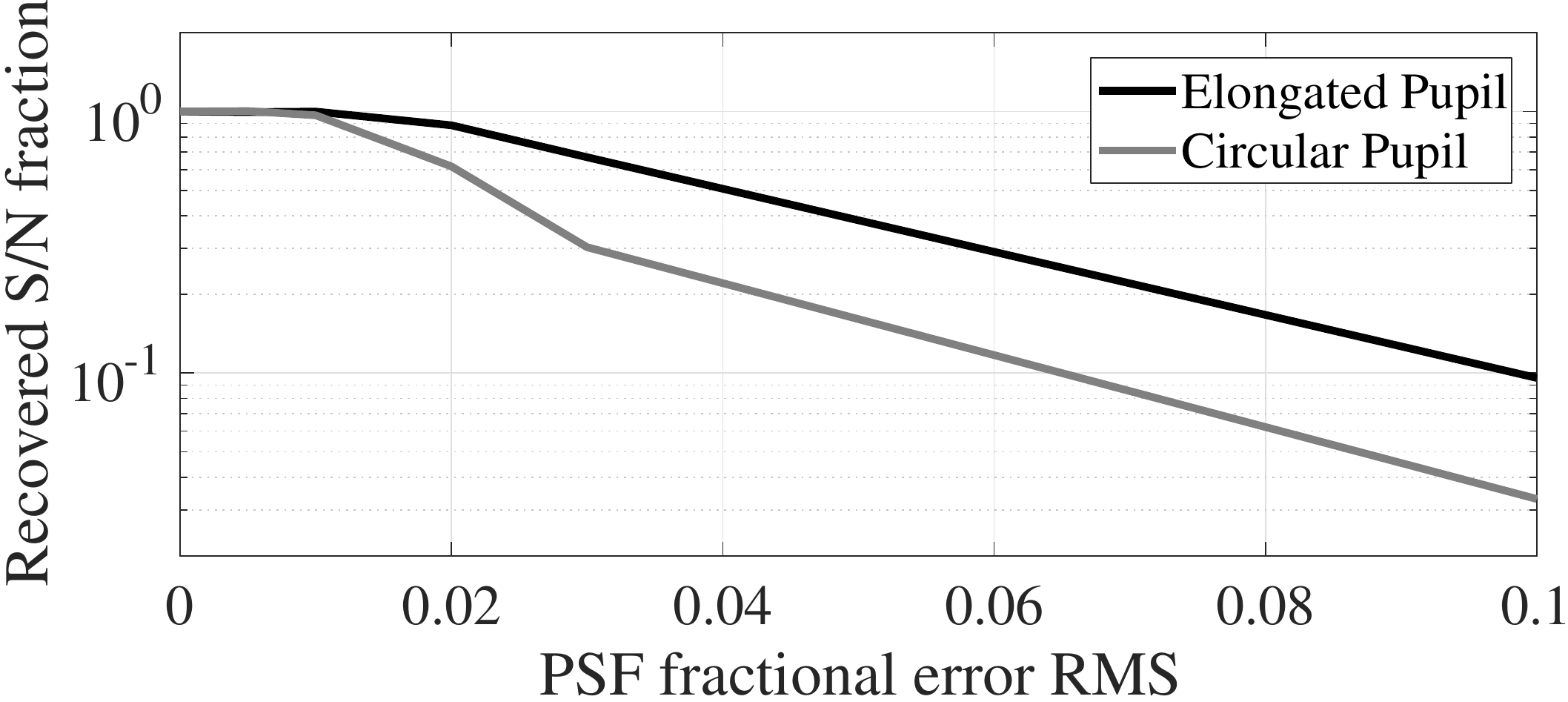}
		\caption{The relative loss of signal caused by fractional PSF errors. 
			The black line shows the loss of signal for the elongated-pupil telescope simulation, 
			while the gray line is for the circular-pupil simulation. 
			The $x$-axis shows the noise RMS of the PSF pixels, 
			given as a fraction of each PSF pixel value. 
			The $y$-axis shows the measured companion signal in the score map, 
			after subtracting the background score 
			and dividing by the standard deviation of the surrounding score map. 
			The signal begins to deteriorate above 1\% error, and is washed out above 10\%. 
		}
		\label{fig: psf error fractional}
	\end{figure}

    The statistical score maps allow the detection of the companion up to a fractional PSF noise of 0.1. 
    For the case of perfect knowledge of the PSF, 
    the bright star at the center of the score map 
    should be completely removed by the coaddition method.
    However, when PSF errors are added, it leaks back into the score map and contaminates the companion's signal. 
    Even though the companion's signal shows a peak at the correct location, 
    the signal is washed out by the false signal from the main star, 
    as seen in the example in Figure~\ref{fig: examples psf errors frac}. 
    To estimate the recovered signal, we measured the intensity of the peak, 
    subtracted from it the average of two points, at a distance of 5 pixels to the left and right of the companion peak, 
    then divided by the standard deviation of all pixels at a distance between 4 and 10 pixels from the companion peak. 
    The resulting loss of signal is evident in Figure~\ref{fig: psf error fractional}. 
    
	Further work on testing the effects of other kinds of PSF inaccuracies 
	will be detailed in~\cite{Coaddition4_Zackay_2018}. 
	Initial tests show that PSFs with red noise optical aberrations 
	suffer even more leakage from the central star. 
	It may be possible to mitigate this effect by aligning the PSF to the image. 
	
	\subsection{The effect of the rotation angle sampling}\label{subsec: rotation angles}
	
\begin{figure}
		\centering
		\pic[\plotsize]{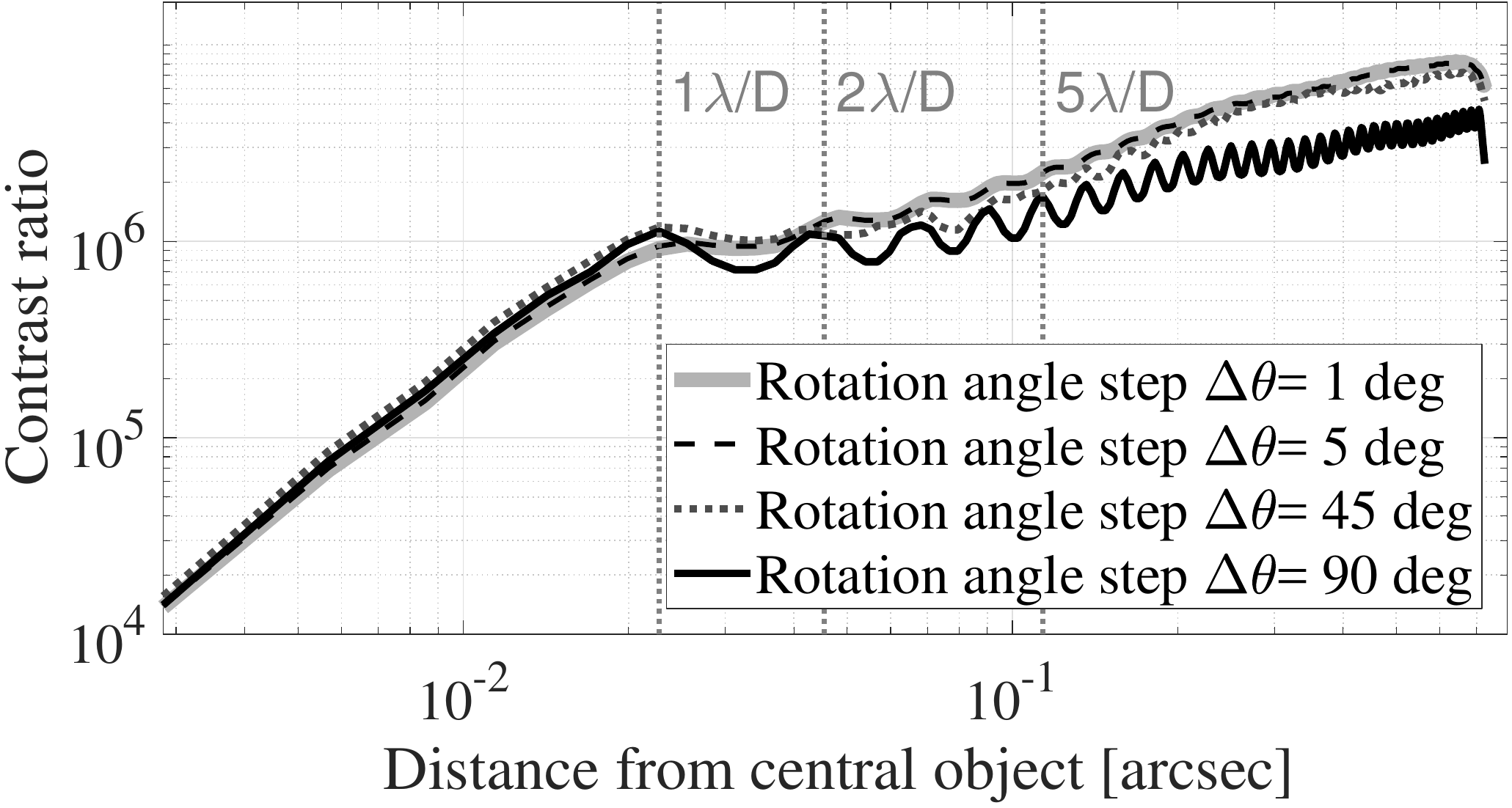}
		\caption{Minimal contrast as a function of angular separation 
			     for several rotation angle step values. 
			     Simulations were performed using the same parameters as in \S\ref{subsec: simulations perfect}, 
			     but with 180 images for each set. 
			     Each simulation used a different value for the rotation angle step $\Delta\theta$. 
			     The minimal contrast to detect a companion in each annulus is plotted, 
			     to highlight the angle step where some parts of the image lose contrast. 
			     We find that for a 1:10 pupil axes ratio, 
			     the contrast is the same up to steps of $\Delta\theta=45$ degrees. 
			     However, by inspection of the coaddition results, 
			     the overall coadded image quality starts to degrade at around $\Delta\theta=5$ degrees. 
			    }
    	\label{fig: angles perfect}
	\end{figure}

	We measured the maximum rotation angle sampling step size 
	that can be used without losing contrast.
	We simulated images using perfect optics, perfect knowledge of the PSF and no atmosphere, 
	and coadded the images as described in previous subsections. 
	As before, we used a $500\times50$\,cm pupil. 
	The simulations were conducted several times using different rotation angle sampling. 
	We used $\Delta \theta=1,5,15,45,60$ and $90$ degree sampling steps.
	In each case, we simulated 180 images. 
	For $\Delta\theta=1$ degree, 
	each image was simulated at a different rotation angle, 
	and for the other simulations, 
	the same angles were repeated several times so that the total number of images was the same. 
	Source noise and an additional 1 electron/pixel noise were added as before.
	The contrast curves remained constant for all rotation angle steps, 
	but at higher step values the 2D contrast surface is less smooth, 
	with polygon modulations dominating the image. 
	Since the contrast image is no longer uniform, 
	we do not expect the median on each annulus to be a good representation of the contrast.
	Instead, we display the minimal contrast in each annulus, 
	which shows the largest loss of contrast when using large rotation angle steps. 	
	The results are shown in Figure~\ref{fig: angles perfect}.
	
	We see that for a 1:10 ratio elongated-pupil telescope, 
	contrast is fairly well maintained up to $\sim 45$ degrees. 
	We also planted a companion with a contrast of $5\times 10^6$, 
	at a separation of $0.3''$, which should be detectable at that distance.
	For all simulations up to and including $\Delta \theta=60$ degrees, 
	the companion is recovered with the expected $S/N$. 
	To calculate a proper image, rather than look for point sources with high contrast ratios, 
	we coadded the simulated images using the proper coaddition technique given by Equation~\ref{eq: proper coaddition full}.
	In the coadded image, we see diffraction spikes around the PSF even at $\Delta \theta=5$ degrees, 
	with increasing intensity of spikes as $\Delta\theta$ gets bigger. 
	It is hard to quantify the effect on image quality from these results. 
	However, we can estimate that the PSF will be sufficiently sharp 
	for angle steps smaller than $\Delta\theta=5$ degrees. 

		
	
	\section{Proof of concept using real measurements}\label{sec: measurements}

	We performed some observations under atmospheric conditions using a small telescope 
	with an elongated pupil. 
	We demonstrate the coaddition method in the case of an unknown PSF and show
	that the basic reconstruction methodology works. 
	
	Observations were taken at the Weizmann Institute 
	during one night in 2018 January 8, 
	using the Kraar observatory 40\,cm telescope at f/25. 
	A fast sCMOS camera (Andor Zyla 5.5) 
	was used to capture speckle images at 100\,Hz with 0.5\,ms exposure times and $\cong 2.5$\,e$^{-}$/pix read noise. 
	The telescope was outfitted with a cardboard mask to simulate a long and narrow aperture. 
	The width of the mask was 2.2\,cm, 
	while its length was 40\,cm, 
	excluding an obscuration of $\cong 10$\, cm in the center of the aperture, 
	where the secondary mirror blocks the light. 
	Observations were made in batches of 3000 images with the mask set at the same orientation during each batch.
	Between each batch, the mask was rotated by $\approx 12$\,deg. 
	We observed Sirius ($M\sim -1.4$) using a V filter. 
	
	We also produced simulations using the same aperture and assuming a Fried length of $r_0=4$\,cm. 
	We matched the total photon count and background noise of the simulations to the observed images.
	We used the same ratio of long-to-narrow aperture, and excluded the central obscuration. 
	The PSF was calculated by generating a random phase screen 
	and then shifting it three times and averaging the resulting PSFs, to simulate a 10\,m/s wind. 
	To match the telescope results, 
	we added specific optical aberrations to the wavefront, 
	in the form of defocus and astigmatism terms. 
	The two terms are given by:
	\begin{equation}\label{eq: defocus astigmatism}
		\Delta\varphi_\text{def}(\rho,\theta)=\sqrt{3}(2\rho^2-1) ,\ \Delta\varphi_\text{ast}(\rho,\theta)=\sqrt{6}\rho^2\cos 2\theta, 
	\end{equation}
	where $\rho$ and $\theta$ are the pupil position parameters, 
	and $\Delta\varphi_\text{def}$ and $\Delta\varphi_\text{ast}$ are, respectively, 
	the defocus and astigmatism added phase term, given in radians.
	The amplitude of each of these aberrations was chosen empirically to be one,
	by comparing simulations to the measured results. 
	On top of these errors, 
	we also added red noise optical aberrations with an amplitude of $A=0.2$, 
	as in Equation~\ref{eq: red noise power law}. 
	Unlike in previous simulations, 
	the optical aberrations were stationary when the elongated pupil mask 
	was rotated during the simulation. 
	This is because in the actual measurements
	we did not rotate the optics of the telescope but only the cardboard mask. 

\begin{figure}
		\centering
		\pic[\plotsize]{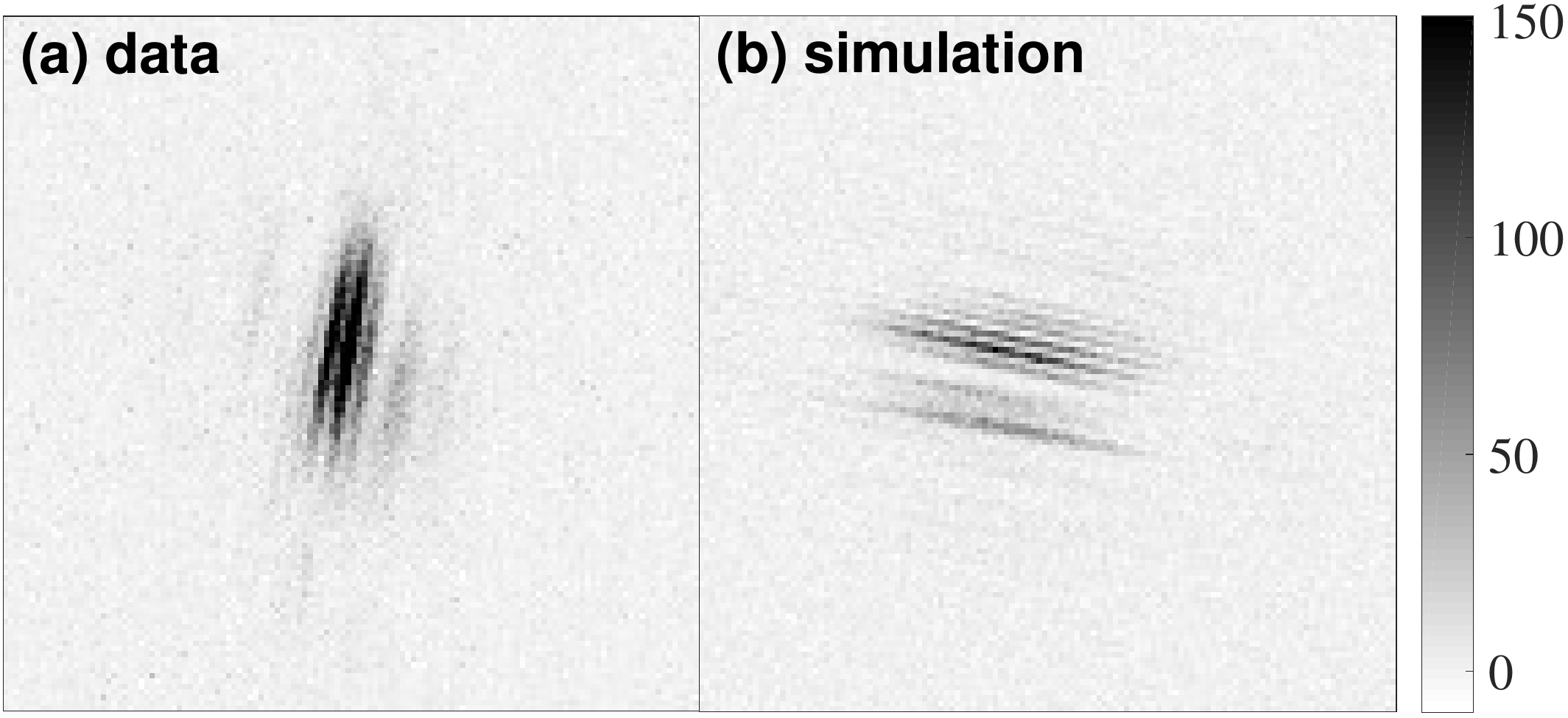}
		\caption{Example images from (a) the Kraar observatory 
			and (b) the simulation following the estimated observational parameters of the data. 
			In these specific examples the data is brighter but the simulation's speckles are sharper. 
			The simulation parameters were chosen so that 
			the average properties of all images, 
			when coadded, would best simulate the coaddition results of the real data. 
		}
		\label{fig: kraar example images}		
	\end{figure}

\begin{figure}
		\centering
		\pic[\plotsize]{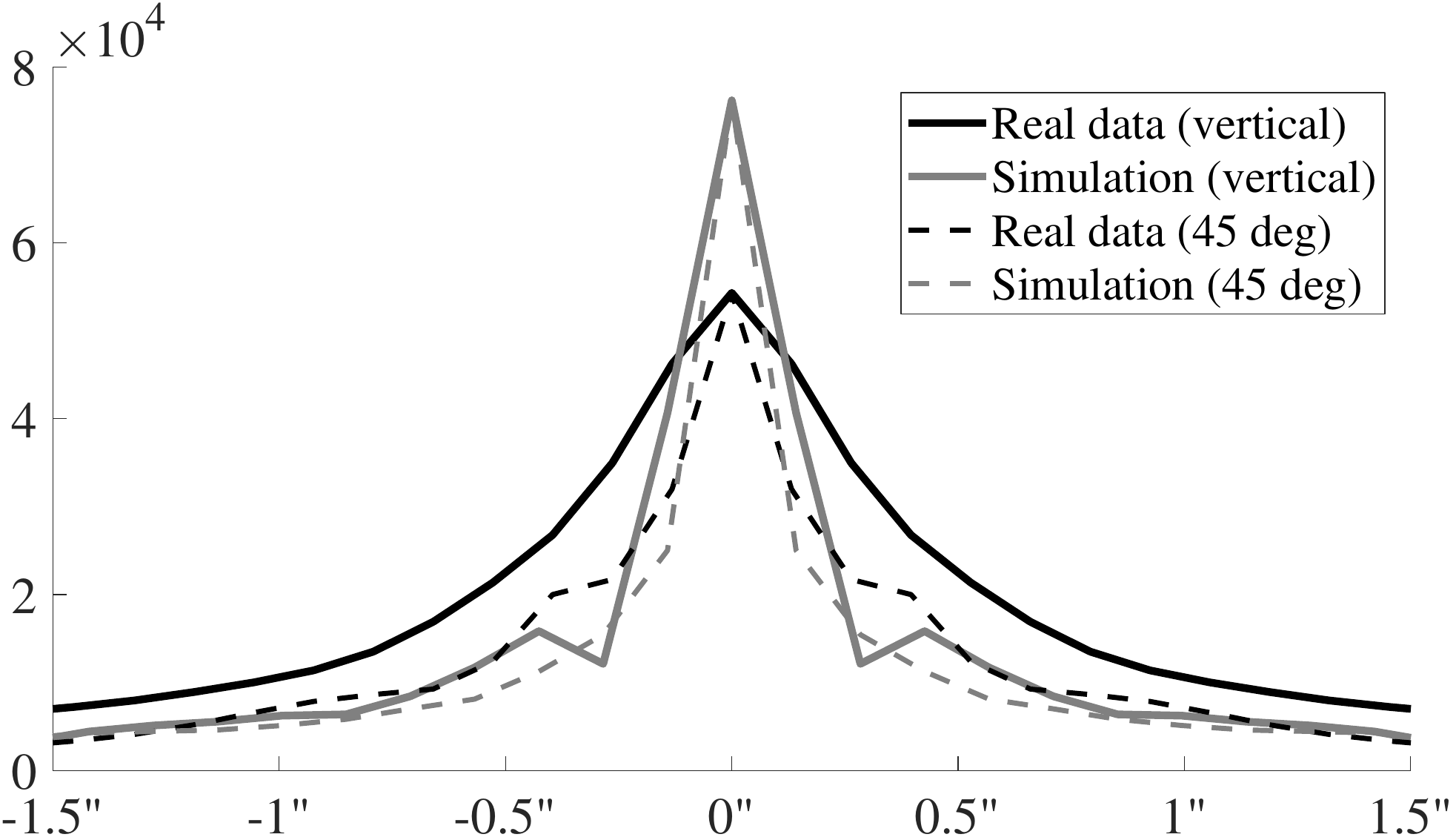}
		\caption{Profiles of the coaddition results of data and simulation.
			The black lines denote real data and the gray lines simulation. 
			The solid line represents a vertical cut through the coaddition result 
			while the dashed line represents a cut through 45 degrees. 
			The simulation coaddition result is still somewhat sharper, 
			indicating some other aberrations (like, e.g., atmospheric color refraction)
			should still be added to perfectly simulate	the telescope used. 
		}
		\label{fig: profiles kraar}		
	\end{figure}

	Some sample images from the simulation and from the actual measurements are shown
	in Figure~\ref{fig: kraar example images}. 
	The effects of the seeing and optical aberrations is visible, 
	though for these specific images the simulation appears to have somewhat sharper speckles. 
	The coaddition result, using Equation~\ref{eq: cosqrt coaddition}, 
	also shows the simulation and data are fairly similar, 
	even though the simulation is a bit sharper. 
	A profile through the coaddition result is shown on Figure~\ref{fig: profiles kraar} 
	for slices through the 2D map, in a vertical direction and in a 45 degree angle. 
		
\begin{figure}
		\centering
		\pic[\plotsize]{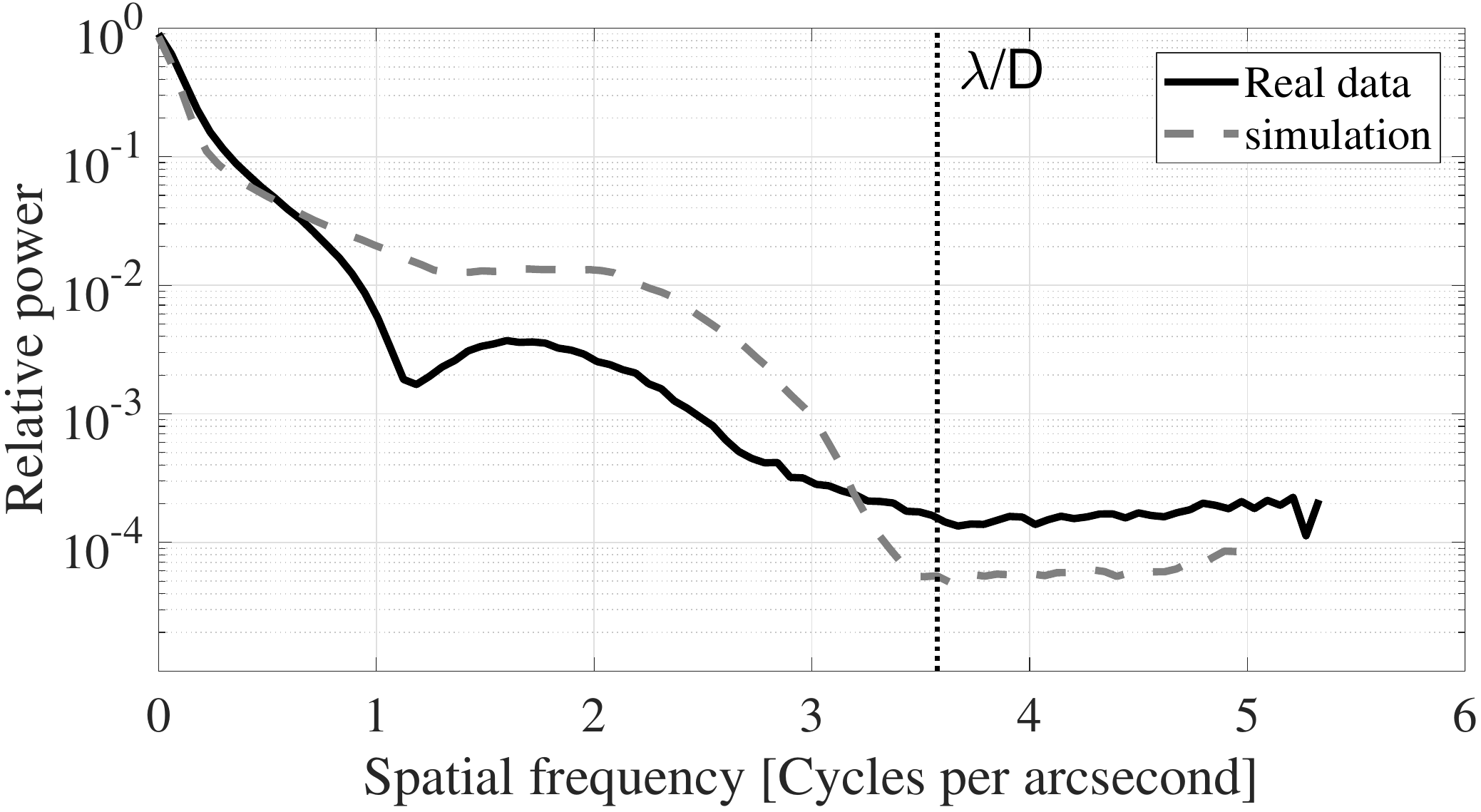}
		\caption{The MTF for real data and simulation, averaging over annuli 
			with various radii around the center of the coaddition result. 
			The black line represents the real data MTF, 
			while the gray line represents the simulation MTF. 
			Once more, the observations still show some unmodeled aberrations 
			as compared to the simulations, but the general features are similar: 
			Both show a drop in the MTF below 1 cycle/arcsecond (above the seeing limit)
			and both start losing information below the diffraction limit due to strong optical aberrations. 			
		}
		\label{fig: mtf kraar}
	\end{figure}

\begin{figure}
		\centering
		\pic[\plotsize]{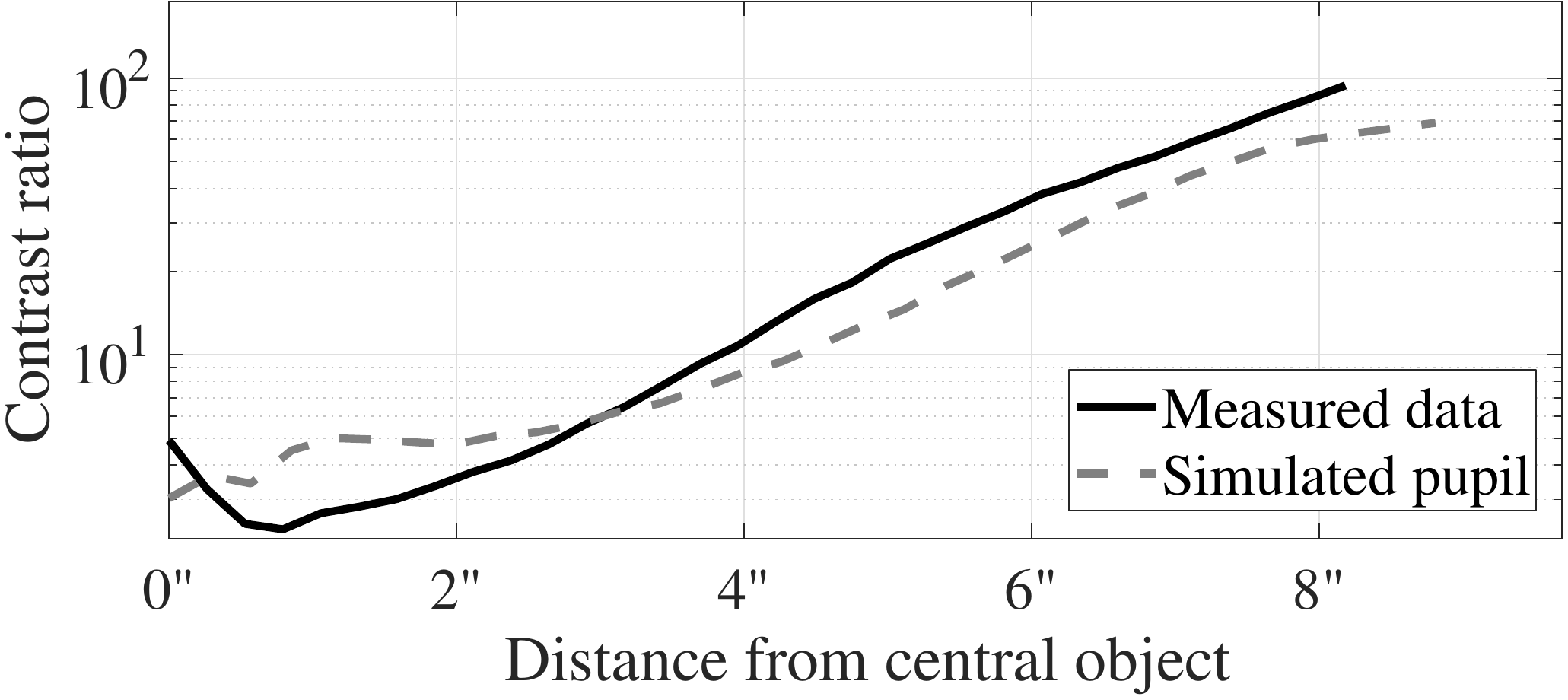}
		\caption{The maximum contrast between a primary star and a companion that can be detected at $S/N=5$ 
			     for simulation and real observations using the Kraar 40\,cm telescope at the Weizmann Institute. 
			     The simulations used the same photon count and background noise as the real measurements. 
			     PSFs were calculated using a $40\times 2.2$\,cm aperture and a Fried length of $r_0=4$\,cm, 
			     corresponding to a seeing of $\approx 2.5$'', similar to the local seeing conditions, 
			     with wind, red noise optical aberrations, defocus and astigmatism added 
			     so the simulation better represents the telescope used. 
			    }
		\label{fig: contrast kraar}
	\end{figure}

	We show the MTFs for the simulation and the data in Figure~\ref{fig: mtf kraar}. 
	We see that the optical aberrations cause a loss of information 
	at scales lower than the diffraction limit. 
	The coaddition can also be used to find the maximal contrast for a companion 
	to be detectable at a given angular separation. 
	The contrast curve results for the simulation and the data are shown in Figure~\ref{fig: contrast kraar}.
	
\begin{figure}
		\centering
		\pic[\plotsize]{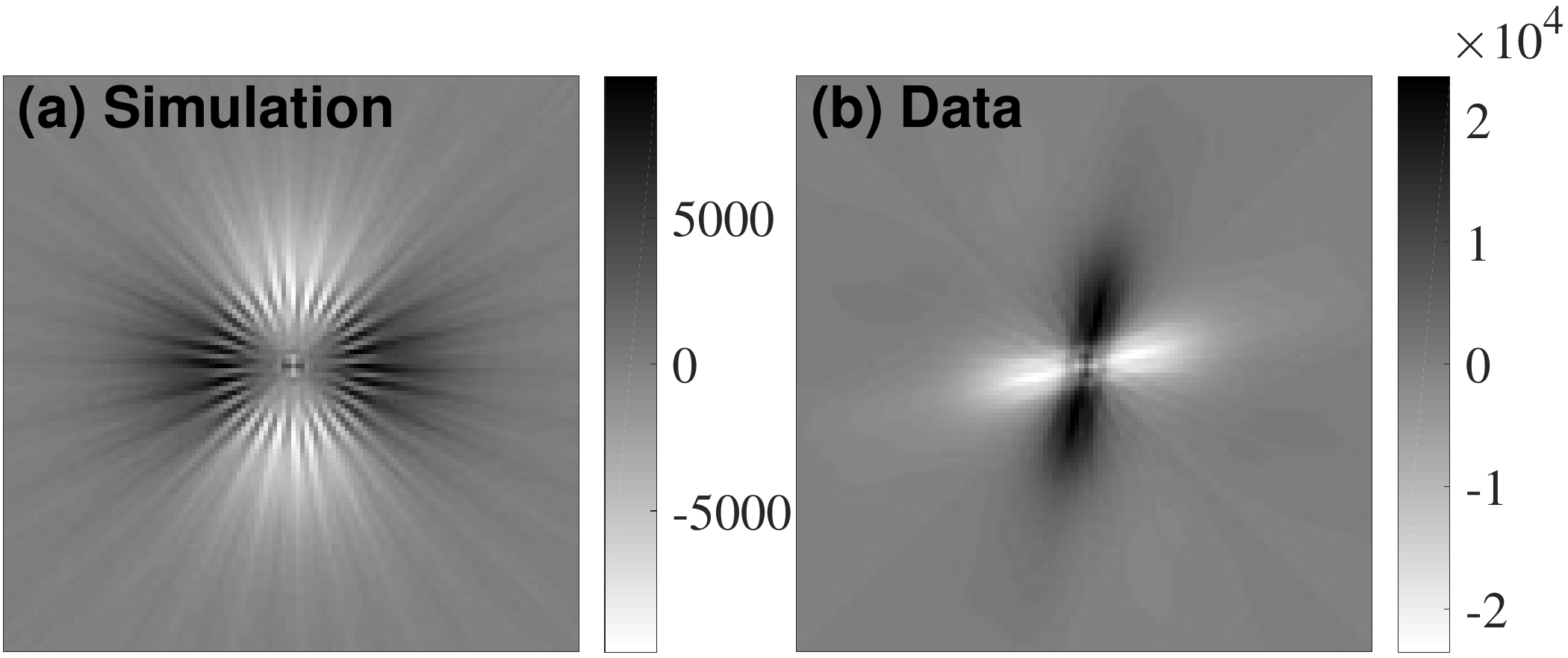}
		\caption{The coaddition results for data and simulation, 
			using symmetric self subtraction, 
			i.e., subtracting the coaddition image from its own transposed image (see, e.g., \citealt{Coaddition2_Zackay_2017})
			Both examples show large asymmetry, which strongly deteriorates the contrast 
			for detection of faint companions. 
			Without an astigmatism term added deliberately to the simulations, 
			we could not recover such asymmetry. 
			Note that in the measurements only the narrow slit mask was rotated, 
			not the whole telescope. 
			If the whole telescope were rotated, this asymmetry would be averaged out over all angles. 
		}
		\label{fig: asymmetry kraar}
	\end{figure}

	The drop in contrast below the seeing limit, 
	which is more pronounced in the data and the simulations discussed in this section 
	relative to the simulations shown in \S\ref{subsec: simulations atmosphere}, 
	is most likely due to the astigmatism of the telescope.
	The asymmetry is hard to notice in individual speckle images, 
	but is apparent in the coaddition result. 
	It is clearly visible in Figure~\ref{fig: asymmetry kraar}, 
	where the image has been subtracted from its own transpose. 
	We calculated the contrast as the ratio of peak intensity to 
	the noise standard deviation in annuli around the peak. 
	The astigmatism asymmetry has a strong effect on the measured noise in each annulus, 
	which reduces the contrast dramatically. 
	It is likely that this effect would be smaller if the astigmatism in the telescope 
	were lower; if the mirror were rotated and the asymmetry averaged out over all angles; 
	if the PSF were measured and used in the coaddition; or, ideally, all of the above. 
	
	\section{Summary}\label{sec: summary}
	
	In this work, we tested the concept of an elongated-pupil telescope 
	and showed that it is able to detect faint companions around bright stars at a 
	higher contrast ratio and closer separation 
	compared with a circular-pupil telescope of equal area. 
	We conducted simulations comparing the two designs 
	using two coaddition algorithms used when the PSF is known and when it is unknown. 
	We tested them on images simulated assuming no atmosphere and under the assumptions 
	of perfect, and then imperfect, red-noise optics, 
	while also assuming perfect knowledge of the PSF.
%
	We performed tests on different rotation angle steps 
	and found the minimal angle needed to maintain contrast for a 1:10 axes ratio. 
	
	We conducted simulations assuming atmospheric conditions 
	in the two cases of known and unknown PSF. 
	In the latter, we applied the coaddition algorithm discussed in \S\ref{subsec: cosqrt}. 
	We also presented observations with a small telescope fitted with a mask limiting the aperture to a long slit,  
	and compared the results to simulations. 
	
	The simulations comparing circular- and elongated-pupil telescopes all show
	that the elongated-pupil telescope has an advantage when searching for high-contrast companions at close separations. 
	We demonstrated that using image coaddition techniques allows the use of new asymmetric designs where higher resolution can be recovered
	while maintaining the total area of the telescope. 
	These results also highlight the importance of using proper image coaddition when adding images with very different PSFs. 
	
	\pagebreak
	
	\section*{Acknowledgments}
	
	G.N.~would like to thank N.~Eliyahu and H.~Azzam for their participation in observations for this work. 
	B.Z.~acknowledges the support from the infosys fund.
	E.O.O.~is grateful for the support by grants from the 
	Israel Science Foundation, Minerva, Israeli Ministry of Science, Weizmann-UK, 
	the US-Israel Binational Science Foundation,
	and the I-CORE Program of the Planning
	and Budgeting Committee and The Israel Science Foundation.
	
	
	\bibliographystyle{aasjournal}
	
	\bibliography{references}
	
\end{document}